\documentclass[10pt, a4paper, preprint]{article}
\usepackage[utf8]{inputenc}
\usepackage{multirow}
\usepackage[T1]{fontenc}
\usepackage{threeparttable} 
\usepackage{graphicx}
\usepackage{authblk}
\usepackage[percent]{overpic}
\usepackage{amsmath,amssymb,amsfonts}

\title{Imaging performance of a Single-Plane Readout Compton Camera\footnote{© 2026 The Authors. This work is licensed under a Creative Commons Attribution 4.0 License. This is the authors’ accepted manuscript of an article published in IEEE Transactions on Nuclear Science, DOI: 10.1109/TNS.2026.3744784.}}

\author{Om Prakash Dash, Tomislav Bokulić, Damir Bosnar,\\ and Mihael Makek\footnote{Correspondence: makek@phy.pmf.unizg.hr}}%

\affil{University of Zagreb Faculty of Science, Department of Physics, Zagreb, 10000, Croatia}

\date{}

\begin{document}

\maketitle

\begin{abstract}
This work presents the design, development, and characterization of a compact Compton camera employing a novel single-plane readout architecture. The proposed concept simplifies conventional multilayer Compton cameras by optically coupling the scatterer and absorber into a single compact detector element using a light guide, enabling readout exclusively from one side with silicon photomultipliers. The system was validated through Monte Carlo simulations, prototype construction, and experimental characterization using standard gamma-ray sources. Detector performance for 511~keV and 662~keV photons was optimized using simulations based on the Geant4 framework, by guiding event selection strategy. Event reconstruction relied on coincidence detection consistent with Compton scattering kinematics, while image reconstruction was performed using maximum likelihood expectation maximization algorithm. The prototype comprises 64 detector elements arranged in an $8\times8$ matrix, read out by an $8\times8$ silicon photomultiplier array. Each element consists of two GAGG:Ce scintillators coupled by a 20 mm long light guide. Measured energy resolutions were $8.9\%\pm1.9\%$ and $10.8\%\pm1.6\%$ for the front and back layers, respectively. Successful gamma-ray imaging was demonstrated for Na-22 and Cs-137 sources, achieving angular resolutions of $12.4^\circ$--$14.3^\circ$ at 511~keV and $14.3^\circ$--$16.8^\circ$ at 662~keV, confirming the feasibility of the proposed single-plane architecture for compact and cost-effective gamma-ray imaging applications.
\end{abstract}

\section{Introduction}
\label{sec:introduction}

Gamma rays, located at the high-energy end of the electromagnetic spectrum, possess strong penetrating power and carry unique information about radioactive sources and physics processes. Their interaction with matter enables a wide range of imaging and spectroscopic techniques that are essential in applications such as nuclear medicine, environmental monitoring, homeland security, and astrophysics. In particular, gamma-ray imaging plays a critical role in medical diagnostics through techniques such as Positron Emission Tomography (PET) and Single-photon Emission Computed Tomography (SPECT), as well as in the detection and localization of radioactive materials in environmental and security-related scenarios.

Conventional gamma-ray imaging systems, such as Anger cameras, rely on mechanical collimators to determine the direction of incoming photons~\cite{anger1958scintillation}. While effective, this approach severely limits detection efficiency and sensitivity due to photon absorption by the collimator. Compton cameras aim to overcome these limitations by exploiting the kinematics of Compton scattering~\cite{compton1923quantum} to perform electronic collimation, enabling wider field-of-view, and improved sensitivity. Since their first proposal by Schönfelder et al.~\cite{schonfelder1973telescope} in astrophysics, and in medical imaging by Todd et al.~\cite{todd1974proposed}, Compton cameras have evolved significantly and are now recognized as a versatile imaging devices across multiple disciplines.

In a typical Compton camera (CC) configuration, an incident gamma photon undergoes Compton scattering in a scatterer detector and may subsequently be absorbed in an absorber detector. By measuring the deposited energies and interaction positions for events in which the scattered photon is fully absorbed, the scattering angle can be determined using Compton kinematics, constraining each event to a conical surface in space. The accumulation of multiple such events enables reconstruction of the gamma-ray source distribution. Early CC implementations were primarily based on semiconductor detectors, which provide excellent spatial resolution but suffer from limited detection efficiency, high system complexity, and elevated cost~\cite{watanabe2007development,takahashi2018semiconductor}. To address these limitations, scintillator-based CCs were later developed, initially coupled with photomultiplier tubes (PMTs)~\cite{tanimori2004mev} and more recently with silicon photomultipliers (SiPMs), allowing reduced system size, lower operating voltage, and improved mechanical robustness~\cite{llosa2019sipm,parajuli2022development}. Although scintillator-based systems typically exhibit lower intrinsic energy resolution compared to semiconductor detectors, they offer significantly higher detection efficiency, reduced cost, and simpler system integration, making them highly suitable for  imaging applications~\cite{jiang2021prototype,zhang2019prototype}.

Building on these advantages, scintillator-based Compton cameras have evolved into several optimized configurations that balance efficiency, spatial resolution, and compactness. Early multi-layer (stacked) designs, such as those demonstrated in COMPTEL, employ discrete scatterer--absorber scintillator layers (e.g., CsI(Tl), LaBr$_3$) to enable efficient multi-interaction event reconstruction and wide-field imaging~\cite{schonfelder1984imaging}. These approaches were further refined through the development of pixelated scintillator arrays using materials such as GAGG:Ce, LYSO, and LaBr$_3$ coupled with SiPMs, providing improved spatial resolution and timing performance, particularly for applications such as real-time hadron therapy monitoring~\cite{kasper2020sifi, barrientos2021performance}. 

In parallel, monolithic scintillator detectors have been introduced to enhance detection efficiency by minimizing dead space while enabling continuous position estimation through light-sharing techniques. For instance, Kataoka et al. developed a lightweight ($\sim$1.5 kg) two-plane Compton camera with depth-of-interaction (DOI) capability, achieving an energy resolution of $\sim$10\% (FWHM) at $^{137}$Cs and improving angular resolution from $14^\circ$ (non-DOI) to $10^\circ$ (DOI), enabling rapid imaging within a minute and successful identification of radioactive isotopes near the Fukushima Daiichi power plant~\cite{kataoka2013handy}. Further improvements in three-dimensional event reconstruction have been achieved using DOI-enabled designs, which enhance angular resolution by resolving the interaction depth within the scintillator volume. 

Recent developments also explore alternative detector technologies, such as the single-layer CdTe Compton camera with Timepix3 readout reported by Turecek \cite{turecek2020timepix3}, featuring a $256 \times 256$ pixel matrix with $55~\mu$m pitch and excellent timing resolution of $1.6$ ns, enabling interaction position estimation via time-of-arrival information and offering potential for intensity mapping and multi-source imaging. Additionally, Barrientos developed the three-layer MACACO II Compton camera for medical imaging, employing two continuous and one monolithic LaBr$_3$ scintillator layers read out by pixelated SiPM arrays, achieving energy resolutions of 5.6--7.2\% (FWHM) at 511 keV and an angular resolution of $8.0^\circ$ (FWHM) at 1275 keV, along with millimeter-scale spatial resolution and measurable coincidence efficiency~\cite{barrientos2021performance}. Recently, the newly developed MACACO III Compton camera, comprising three continuous LaBr$_3$:Ce scintillator planes coupled to SiPM arrays, demonstrated enhanced background rejection and successful clinical imaging of radiopharmaceuticals such as $^{131}$I and $^{18}$F-FDG, highlighting its potential for nuclear medicine applications. An upgraded version, MACACO III+, incorporated a larger active detector area of the second plane to improve sensitivity, enabling the tracking of therapeutic radionuclides including $^{131}$I and $^{225}$Ac in preclinical studies with acquisition times shorter than those of conventional preclinical SPECT systems~\cite{roser2025macacoIII}.

More recently, hybrid scintillator configurations, such as GAGG-based Compton--PET systems, have demonstrated the potential for multimodal imaging by combining high energy resolution with flexible and versatile detector geometries~\cite{shimazoe2020development}.
In parallel, alternative compact architectures based on a single segmented or voxelized scintillator have also been investigated. Lee et al. developed a position-sensitive $4\pi$ Compton camera based on a single segmented scintillator, demonstrating three-dimensional interaction localization and Compton imaging using a compact scintillator-based detector architecture~\cite{Lee2020}. Subsequently, Choi et al. investigated the detection efficiency and imaging performance of a larger single voxelized scintillator Compton camera, further demonstrating the potential of such architectures for improving detector efficiency and maintaining compact system configurations~\cite{Choi2025}.

Despite these advances, conventional multi-plane Compton camera architectures remain limited by mechanical complexity, large channel counts, and increased system thickness, which hinder portability and scalability. To address these challenges, compact detector concepts with reduced readout complexity have gained growing interest. In this context, single-plane Compton camera designs, in which scattering and absorption occur within a unified detector structure, represent a promising alternative.


In this work, we report on the development and characterization of a compact Compton camera implementing a single-plane, single-side readout concept~\cite{dash2023design}. The detector employs pixelated GAGG:Ce scintillators in which the scatterer and absorber elements are optically coupled via light guides and read out by a single SiPM array. By reducing the number of readout channels relative to conventional dual-layer Compton cameras, this design offers a simpler and more compact architecture while preserving the information necessary for Compton image reconstruction. The detector performance is assessed through experimental measurements with 511~keV and 662~keV gamma-ray sources, providing an evaluation of the performance and imaging capabilities of the proposed concept.

\section{Principle of Operation}
\label{sec:principle}

Compton cameras exploit the kinematics of Compton scattering to estimate the direction of incident gamma rays without mechanical collimation. Directional information is inferred from the measured interaction positions and deposited energies, enabling significantly higher detection efficiency compared to collimator-based gamma cameras.

\subsection{Conventional Two-Plane Compton Camera}

\begin{figure}
\centering
\includegraphics[width=1\linewidth]{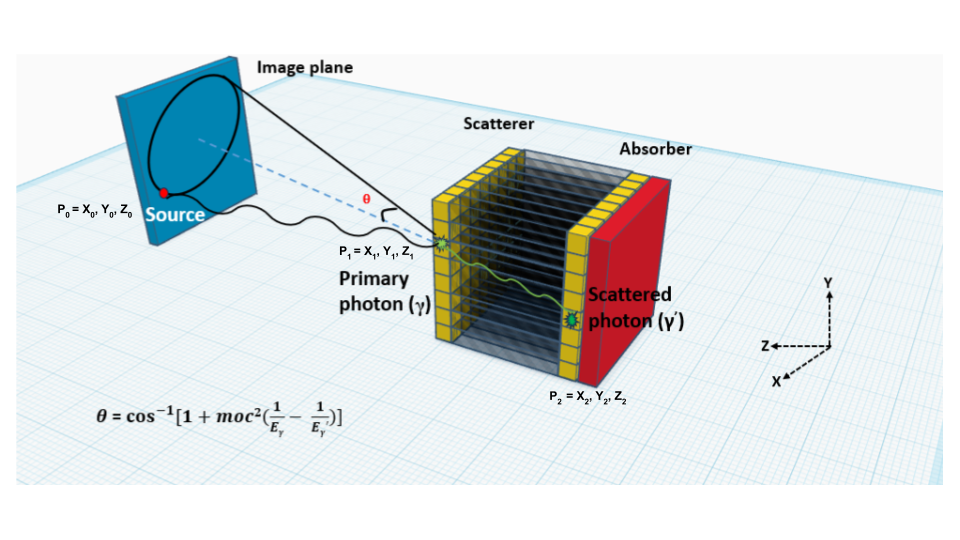}
\caption{Schematic of a conventional two-plane Compton camera illustrating Compton scattering in the scatterer and full absorption in the absorber.}
\label{fig:two_plane_cc}
\end{figure}

In the conventional configuration, a Compton camera consists of two spatially separated detector planes: a scatterer followed by an absorber. As illustrated in Fig.~\ref{fig:two_plane_cc}, an incident gamma photon undergoes Compton scattering in the first detector, depositing part of its energy and producing a recoil electron. The scattered photon then propagates toward the second detector, where it may be fully absorbed through the photoelectric effect.

Only events satisfying this interaction sequence are used for image reconstruction. The separation between the two planes is typically optimized to balance angular resolution and intrinsic detection efficiency, as the angular uncertainty is inversely proportional to the distance between scatterer and absorber, while the efficiency is inversely proportional to the square of the distance~\cite{du2001evaluation}. The scattering angle $\theta$ is determined from the deposited energies using Compton kinematics. For an incident photon of energy $E_\gamma$ scattered to energy $E_{\gamma'}$, the angle is given by
\begin{equation}
\cos\theta = 1 - \frac{m_e c^2 (E_\gamma - E_{\gamma'})}{E_\gamma E_{\gamma'}},
\end{equation}
where $m_e$ is the electron rest mass and $c$ is the speed of light.
Expressed in terms of the recoil electron energy deposited in the scatterer, $E_e$, the scattering angle becomes
\begin{equation}
\theta = \cos^{-1} \left( 1 - \frac{m_e c^2}{E_\gamma}
\frac{E_e}{E_\gamma - E_e} \right).
\label{eq:theta}
\end{equation}

In the present experiment, the incident photon energy is assumed to be unknown, and therefore it is determined as the sum of the energy deposits of the recoil electron and the scattered photon:
\begin{equation}
E_\gamma = E_e + E_{\gamma'},
\end{equation}


Each valid interaction defines a \emph{Compton cone}, with the vertex corresponding to the scatter interaction position, axis is defined by the line connecting the scatter and absorption points, and half opening angle is given by $\theta$. Image reconstruction is achieved by back-projecting these cones into the image space. In simple back-projection (BP), the reconstructed image is formed by directly back-projecting the measured Compton cones into the image space. Maximum Likelihood Expectation Maximization (MLEM) reconstructs the same image iteratively using a statistical model of the imaging system, where repeated forward projection and back-projection refine the activity distribution, typically resulting in improved localization, contrast, and reduced image artifacts.

\subsection{Performance Parameters}
\label{sec:performance_metrics}

The performance of a Compton camera is primarily characterized by its angular resolution and intrinsic detection efficiency. These metrics quantify the accuracy of the reconstructed Compton cone and the probability of detecting and reconstructing usable events, respectively.

\subsubsection*{Angular Resolution}

The angular resolution describes the uncertainty in the reconstructed angular direction of the gamma source. The total angular uncertainty arises from three independent contributions: detector energy resolution, spatial resolution (finite pixel size and detector separation), and Doppler broadening due to bound electron's momentum. The latter introduces a fundamental lower bound on angular resolution that is material dependent and becomes significant at large scattering angles~\cite{Kishimoto2016study}. The angular resolution is commonly quantified using the Angular Resolution Measure (ARM), defined as the difference between the Compton scattering angle obtained from energy deposition and that derived from detector geometry:
\begin{equation}
\mathrm{ARM} = \theta - \theta_{g},
\label{eq:ARM}
\end{equation}
where $\theta$ is calculated from the measured energy deposits according to Compton kinematics (see Eq.~\ref{eq:theta}), and $\theta_{g}$ is the geometrical scattering angle determined from the known source position $\mathbf{P}_0$ and measured interaction positions $\mathbf{P}_1$ and $\mathbf{P}_2$ as shown in Fig.~\ref{fig:two_plane_cc}:
\begin{equation}
\cos \theta_{g} =
\frac{(\mathbf{P}_1 - \mathbf{P}_0)\cdot(\mathbf{P}_2 - \mathbf{P}_1)}
{|\mathbf{P}_1 - \mathbf{P}_0|\,|\mathbf{P}_2 - \mathbf{P}_1|}.
\label{eq:theta_geom}
\end{equation}

The ARM distribution is centered around zero for correctly reconstructed events, and its full width at half maximum (FWHM) is used as a measure of the system angular resolution. 




\subsubsection*{Intrinsic Detection Efficiency}

The intrinsic efficiency quantifies the probability that an incident gamma ray interacting with the detector contributes to a valid, reconstructable Compton event. It is defined as the ratio of detected events to the number of incident photons on the detector volume~\cite{leo1994techniques}:
\begin{equation}
\epsilon_{\mathrm{int}} =
\frac{N_{\mathrm{detected}}}{N_{\mathrm{incident}}}.
\label{eq:intrinsic_eff_def}
\end{equation}

In this work, the intrinsic efficiency is estimated by combining experimentally measured reconstruction efficiency with simulation-based detection probability. The probability that a detected event satisfies reconstruction criteria is given by:
\begin{equation}
P_{\mathrm{recon}} =
\frac{N_{\mathrm{recon}}^{\mathrm{exp}}}
{N_{\mathrm{detected}}^{\mathrm{exp}}},
\label{eq:precon}
\end{equation}
while the probability of detecting an incident gamma ray is obtained from Monte Carlo simulations:
\begin{equation}
P_{\mathrm{det}} =
\frac{N_{\mathrm{detected}}^{\mathrm{sim}}}
{N_{\mathrm{incident}}^{\mathrm{sim}}}.
\label{eq:pdet}
\end{equation}

The intrinsic efficiency of the Compton camera is then expressed as:
\begin{equation}
\epsilon_{\mathrm{intrinsic}} =
P_{\mathrm{recon}} \times P_{\mathrm{det}} =
\frac{N_{\mathrm{recon}}^{\mathrm{exp}}}
{N_{\mathrm{detected}}^{\mathrm{exp}}}
\cdot
\frac{N_{\mathrm{detected}}^{\mathrm{sim}}}
{N_{\mathrm{incident}}^{\mathrm{sim}}}.
\label{eq:intrinsic_efficiency}
\end{equation}
where exactly the same selection criteria are applied to determine the $N_{\mathrm{detected}}^{\mathrm{exp}}$ and $N_{\mathrm{detected}}^{\mathrm{sim}}$. This formulation provides a robust and experimentally meaningful measure of system performance by jointly accounting for detector response and reconstruction capability.

\subsection{Single-plane readout Compton camera concept}

While dual plane Compton cameras provide robust imaging performance, their practical deployment is often limited by detector size, mechanical alignment requirements, and readout complexity. To address these limitations, a single-plane readout Compton camera concept was introduced~\cite{dash2023design, dash2026thesis}.

\begin{figure}
\centering
\includegraphics[width=0.75\linewidth]{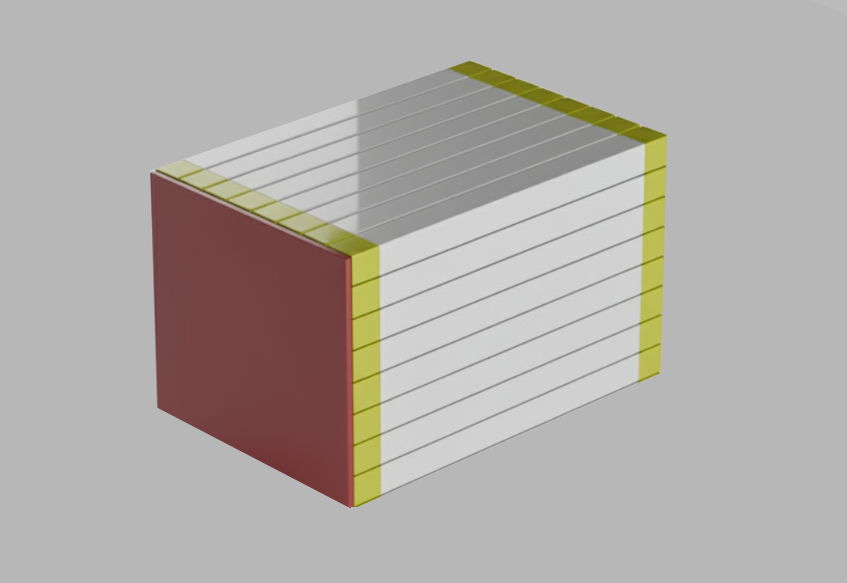}
\caption{Conceptual schematic of the single-plane Compton camera with single-side readout, where scatterer and absorber are optically coupled via a light guide.}
\label{fig:single_plane_cc}
\end{figure}

In this architecture, the scatterer and absorber elements are optically coupled using light guides to form a compact, unified detector plane. Both interaction layers are read out from a single side using a common photo-sensor array, eliminating the need for separate readout planes and reducing the number of electronic channels. A schematic of the concept is shown in Fig.~\ref{fig:single_plane_cc}.

Despite its compact geometry, the fundamental event reconstruction principle remains unchanged. The interaction sequence is inferred from deposited energy and traversed pixel distance, allowing identification of scatter and absorption events consistent with Compton kinematics (see Section \ref{sec:Data_selection_and_analysis}). The resulting Compton cones are reconstructed using the same formalism as in conventional CC systems and then passed to image reconstruction algorithms.

This work demonstrates the experimental validation of the single-plane readout Compton camera concept, and presents imaging performance with Na-22 and Cs-137 point sources.

\section{Experimental setup}
\label{sec:detector_characterization}

To validate the feasibility of the proposed single-plane, single-side readout Compton camera concept, a series of preliminary laboratory measurements were performed on individual and sandwiched scintillator detector elements. These studies focused on optimizing optical coupling, reflector materials, light guide geometry, SiPM selection and readout performance prior to assembling the full detector matrix. Most importantly, these investigations confirmed experimentally that two scintillator crystals connected via lightguide can be read out at one end~\cite{dash2026thesis}. 

\subsection{Selection of Materials}
\label{subsec:material_selection}

The selection of detector materials and geometry is a key factor governing the performance of a scintillator-based Compton camera, as it directly affects energy resolution, spatial accuracy, detection efficiency, and consequently the angular resolution. Particular attention was therefore given to the choice of scintillator, reflective material, light guide, and photo-sensor to ensure efficient optical coupling and reliable signal reconstruction.

For both the scatterer and absorber elements, GAGG:Ce (Gadolinium Aluminum Gallium Garnet doped with Cerium) scintillators were employed.
GAGG:Ce offers a high light yield up to 57000 photons per 1 MeV gamma energy, high density ($6.6~\mathrm{g/cm^3}$) and effective atomic number (54.4) provide a high gamma-ray interaction probability, which is particularly advantageous for compact detector geometries. In addition, GAGG:Ce is non-hygroscopic, enabling the fabrication of finely pixelated arrays without complex encapsulation, and it exhibits no intrinsic radioactivity, thereby avoiding internal background contributions. These properties make GAGG:Ce a highly suitable scintillator material for Compton imaging applications \cite{kamada2011composition, prusa2013light, yoshikawa2013crystal}.

Optical connection between the scintillator pixels was achieved using a plexiglass (PMMA) light guide. PMMA offers high optical transparency in the visible range and supports efficient photon transport through total internal reflection.

Silicon photomultiplier array from Hamamatsu (S13361-3050AE-08) was chosen owing to the fair compatibility of its quantum efficiency with the GAGG emission spectrum.

To enhance light collection efficiency and prevent optical cross-talk between adjacent detector elements, 3M\texttrademark~Enhanced Specular Reflector (ESR) film was used as the reflective material. ESR was chosen for its high  reflectivity exceeding 98\% over the visible spectrum, and low thickness (65~$\mu$m), important for maintaining element pitch equivalent to the pitch of the SiPM array. 

The refractive indices of the materials involved were as follows: for GAGG:Ce ($n \approx 1.85$), PMMA ($n \approx 1.49$), and SiPM ($n \approx 1.41$–$1.53$). Efficient photon transmission across all optical interfaces was ensured thought coupling by optical cement EJ-500 ($n \approx 1.56$).

\subsection{Detector Matrix Assembly}

A detector matrix consisting of 64 detector elements in $8 \times 8$ configuration was assembled. Each element comprised two $3~\mathrm{mm} \times 3~\mathrm{mm} \times 3~\mathrm{mm}$ GAGG:Ce scintillator coupled by a PMMA light guide $3~\mathrm{mm} \times 3~\mathrm{mm} \times 20~\mathrm{mm}$. 

\begin{figure}[!ht]
    \centering
    \begin{overpic}[width=1\linewidth]{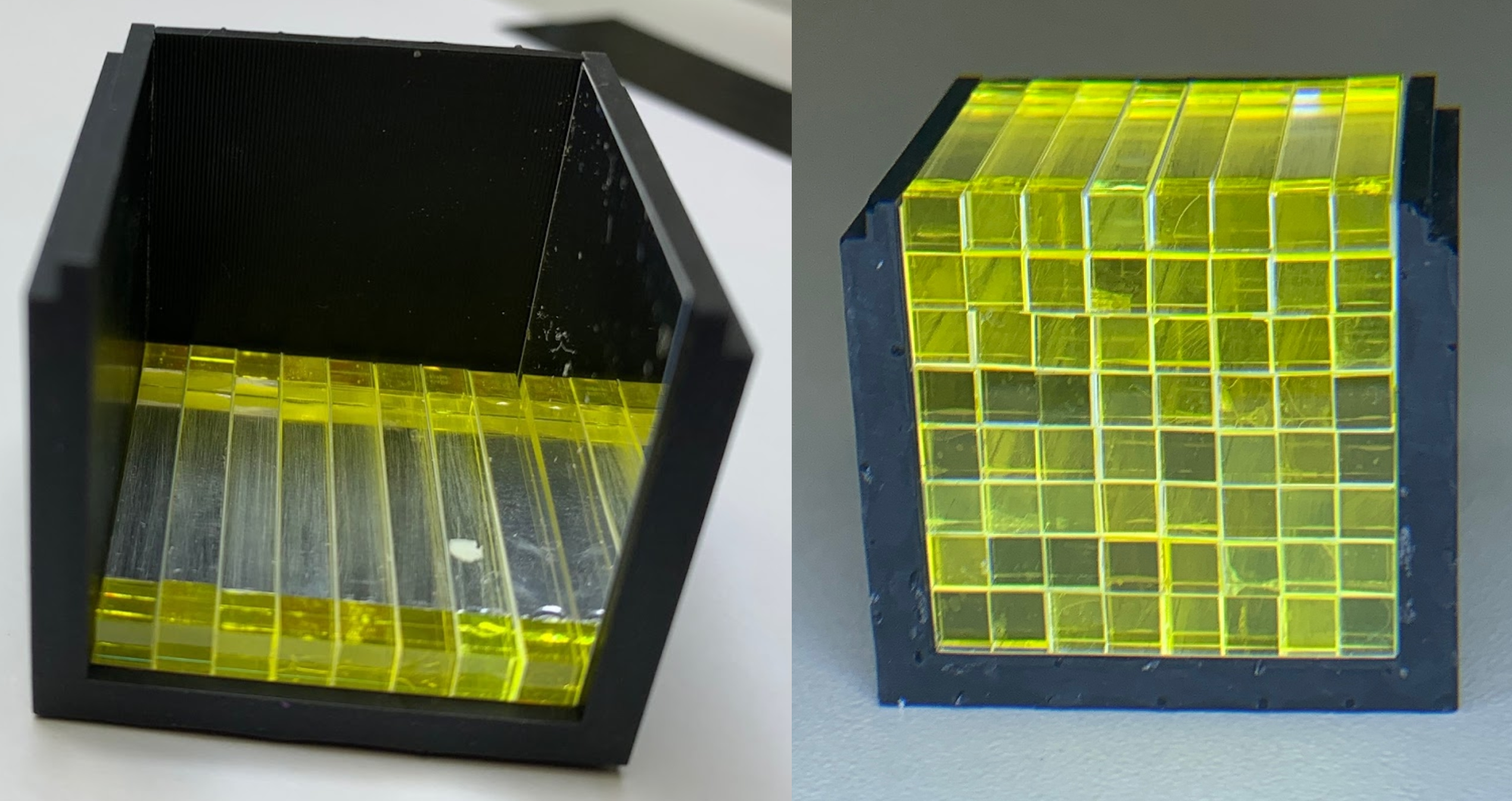}
        \put(1,45){\textbf{(a)}}
        \put(55,45){\textbf{(b)}}
    \end{overpic}
    \caption{(a) Picture of 8 constructed individual elements placed inside the 3D printed opaque housing with ESR materials around each pixel; (b) the complete $8 \times 8$ element matrix within the 3D-printed housing.}
    \label{fig:matrix_assembly}
\end{figure}

Prior to assembly, the light guides were equalized in length using a precision 3D-printed guide and fine sanding, then polished to remove uneven edges and to optimize optical coupling. Each element was then glued using optical cement (EJ-500, Eljen Technology), forming a strong and transparent bond that minimized light loss. One row of the individual elements within a  3D-printed housing is shown in Figure~\ref{fig:matrix_assembly} (a). The housing, was designed to secure pixels with uniform spacing while allowing insertion of ESR foils.

The matrix was assembled row by row. ESR sheets were placed along the inner walls of the housing, followed by placement of eight pixels per row. The 65~$\mu$m-thick ESR sheets were precisely cut and placed between the elements. Vertical strips of $3~\text{mm} \times 26~\text{mm}$ were inserted between adjacent elements, while horizontal sheets of $25.4~\text{mm} \times 26~\text{mm}$ covered the top and bottom surfaces of each row. The process was repeated until the full $8 \times 8$ matrix was completed, and a top ESR sheet was added to enclose the array. Alignment and uniform pitch (3.2~mm) were verified visually, and additional ESR layers (65~$\mu$m) were added selectively to correct minor spacing deviations.

After completing the scintillator matrix, a 64-channel SiPM array (Hamamatsu S13361-3050AE-08) was optically glued to one end of the matrix using EJ-500 optical cement. The SiPM array was then connected to the TOFPET2 \cite{difrancesco2016tofpet2} data acquisition system for testing and characterization.

\subsection{Readout Electronics}

Data acquisition was performed using the TOFPET2 ASIC, a high-performance readout and digitization chip optimized for Silicon Photomultiplier (SiPM) signals. The ASIC provides 64 independent channels, each integrating amplification, discrimination, time-to-amplitude conversion (TAC), charge integration, and digitization~\cite{difrancesco2016tofpet2}. Performance of the GAGG scintillator pixel detectors with TOFPET2 ASIC was previously reported in~\cite{makek2020gagg}.

The SiPM and DAQ parameters used for all measurements are summarized in Table~\ref{tab:sipm_daq_params}. These include bias and overvoltage settings, trigger thresholds, coincidence and digitization windows, and the number of active channels. This configuration ensured high-resolution timing and energy measurements from the detector array, enabling accurate calibration and energy resolution characterization.

\begin{table}
\centering
\caption{TOFPET2 DAQ parameters.}
\label{tab:sipm_daq_params}
\begin{tabular}{|l|c|c|}
\hline
\textbf{Parameter} & \textbf{Value} & \textbf{Notes} \\
\hline
Port ID & 0 & DAQ board port \\
Slave ID & 0 & ASIC slave ID \\
Channels & 0--64 & Active SiPM channels \\
Offset V & 0.75 & Voltage offset per channel \\
Pre-breakdown V & 45.00 & Before SiPM avalanche \\
Breakdown V & 51.40 & SiPM breakdown voltage \\
Overvoltage V & 4.00 & Bias above breakdown \\
Trigger Type & Builtin & Acquisition trigger \\
Trigger Threshold & 1 & Min. photoelectrons \\
Coincidence Window (ns) & 3 & Coincidence timing \\
Pre-window (ns) & 3 & Pre-trigger capture \\
Post-window (ns) & 15 & Post-trigger capture \\
Single Fraction & 0 & Single hit fraction \\
ASIC Disc LSB (ps) & 62 & Discriminator resolution \\
\hline
\end{tabular}
\end{table}

\section{Measurements and Data Processing}
\label{sec:electronics_calibration_analysis}

The detector consists of two GAGG:Ce layers arranged in a sandwiched geometry and read out from one side by a SiPM array. The layer adjacent to the SiPM is referred to as the \textit{back layer}, while the opposite layer is defined as the \textit{front layer}. 
Due to differences in number of optically transmissive and reflective sides, the absorption in the two layers results in slightly different light yields and energy responses. Up to 20\% higher energy response was observed in case of the front layer compared to the back layer crystals \cite{dash2023design}. Corresponding energy resolutions were $8.9\%\pm1.9\%$ and $10.8\%\pm1.6\%$, respectively. We hypothesize, this is due to the front crystal having 5 reflective and 1 transmissive sides, while the crystal adjacent to the SiPM has 4 reflective and 2 transmissive sides, resulting in less light reaching the SiPM from the latter.

\begin{table}[!t]
\caption{Summary of calibration and imaging measurements performed.}
\label{tab:measurement_summary}
\centering
\resizebox{\columnwidth}{!}{%
\begin{tabular}{c l c c c c}
\hline
\textbf{Dataset} & \textbf{Purpose} & \textbf{Source} & \textbf{Position} & \textbf{Distance} & \textbf{Duration} \\
\hline

1 & Low-energy calibration & $^{241}$Am (59.4 keV) & Center (front) & In contact & 24 h \\
2 & Low-energy calibration (partial) & $^{241}$Am & Left / Right sides & Side exposure & 24 h (each) \\
3 & Photopeak calibration & $^{22}$Na (511 keV) & Right (front \& back) & Side exposure & 24 h \\
4 & Photopeak calibration & $^{137}$Cs (662 keV) & Right front (a) \& back (b) & Side exposure & 24 h \\
5 & Photopeak calibration & $^{137}$Cs (662 keV) & Right front (a) \& back (b) & Side exposure & 24 h \\
6 & Imaging (central) & $^{137}$Cs & (0, 0) & 5 cm & 24 h \\
7 & Imaging (central) & $^{137}$Cs & (0, 0) & 10 cm & 24 h \\
8 & Imaging (off-axis) & $^{137}$Cs & (0 mm, $\pm 10$ mm), ($\pm 8$ mm, $\pm 8$ mm) & 10 cm & 24 h (each) \\
9 & Imaging (central) & $^{22}$Na & (0, 0) & 10 cm & 24 h \\
10 & Imaging (off-axis) & $^{22}$Na & (0, 0), (10 mm, 0), (30 mm, 0) & 10 cm & 24 h (each) \\

\hline
\end{tabular}%
}
\end{table}

\begin{figure}[!ht]
    \centering
    \begin{overpic}[width=1\linewidth]{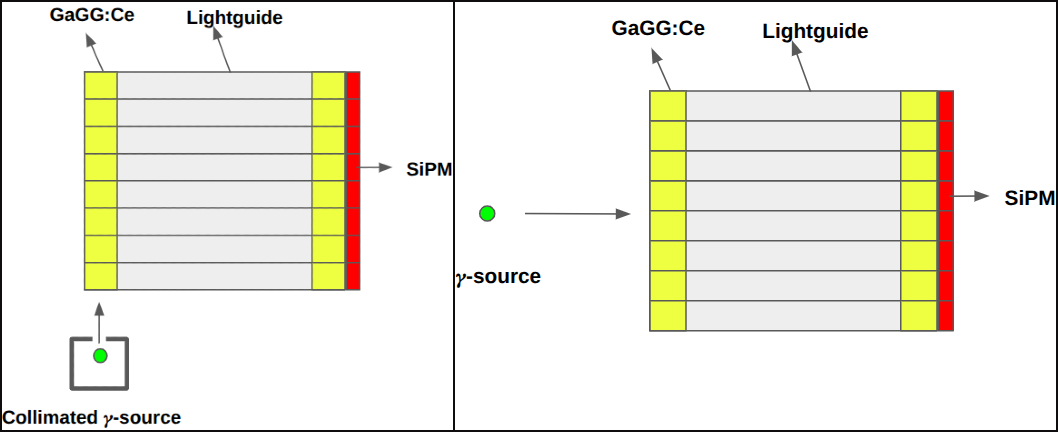}
        \put(38,1){\textbf{(a)}}
        \put(95,1){\textbf{(b)}}
    \end{overpic}
    \caption{Schematic of experimental arrangement for calibration measurements (a), and arrangement for imaging characterization (b).}
\label{fig:Schematic_detector_source}
\end{figure}

Detector characterization and imaging measurements were performed during two 
measurement campaigns. The summary of the performed measurements is given in the table~\ref{tab:measurement_summary}. Typically, two experimental arrangements were set up, the first for the energy calibration, and the second for the imaging characterization measurements, as depicted in Fig. \ref{fig:Schematic_detector_source}.

For all measurements and image reconstructions, the origin of the coordinate system was defined at the centre of the detector's front surface. The \(x\)- and \(y\)-axes represent the lateral directions, while the \(z\)-axis is normal to the detector plane pointing toward the source region. The source activities were 190~kBq for \(^{22}\mathrm{Na}\) and 58~kBq for \(^{137}\mathrm{Cs}\), with an estimated uncertainty of 30\% for both sources. Typically we took 24h measurements, each resulting in $\sim\!1.0\times10^8$ and $\sim\!1.4\times10^7$ acquired events for the $^{22}\mathrm{Na}$ and the $^{137}\mathrm{Cs}$, respectively. After applying the selection criteria (see Section \ref{sec:Data_selection_and_analysis}), the images were reconstructed with $\sim\!2.3\times10^3$ and $\sim\!2.1\times10^3$ events for the two sources, respectively.
All data were acquired at room temperature, typically varying between 21$^\circ$C--23$^\circ$C.

\subsection{Energy Calibration}
The energy calibration was performed for each pixel in each detector element, individually, totaling to 128 separate calibration curves. Calibration measurements were conducted using \textsuperscript{22}Na (511~keV) and \textsuperscript{137}Cs (662~keV) collimated point sources. All sources were placed on the side of the detector, with collimators limiting the gamma-rays to either front or back detector layer. 

To strengthen the calibration accuracy in the low-energy region, dedicated measurements were performed using a \textsuperscript{241}Am source emitting 59.4 keV photons. The source was positioned directly in front of the detector to ensure that these low-energy photons were fully absorbed in the 3~mm thick GAGG:Ce front layer. Under this configuration, all front layer channels could be calibrated using three reference energies. However, due to the placement of the SiPM array immediately behind the scintillator pixels, an equivalent measurement for the back layer was not feasible. When the \textsuperscript{241}Am source was positioned on the left and right sides of the back layer of the detector, a usable 59.4~keV peak could be extracted for only 16 back layer channels.

\begin{figure}[!ht]
    \centering
    
    \begin{overpic}[width=0.8\linewidth]{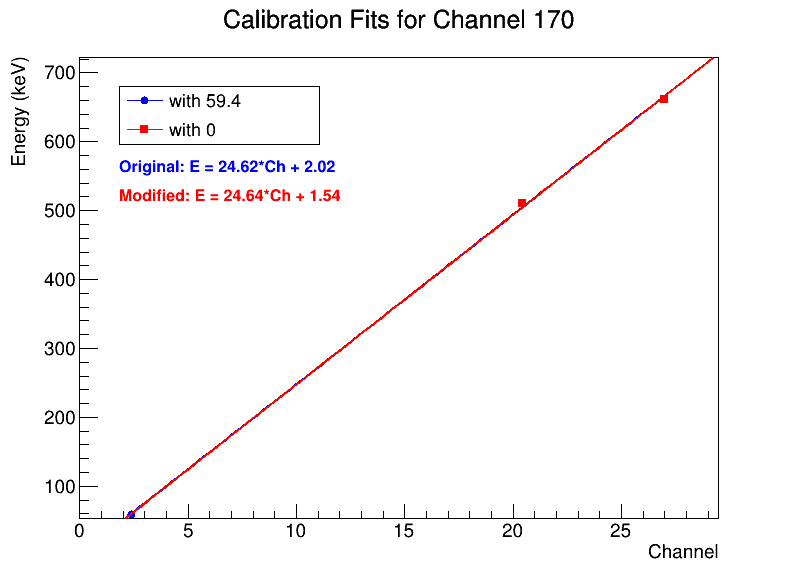}
        \put(80,10){\textbf{(a)}}
    \end{overpic}
    
    \vspace{0.5em}
    
    \begin{overpic}[width=0.8\linewidth]{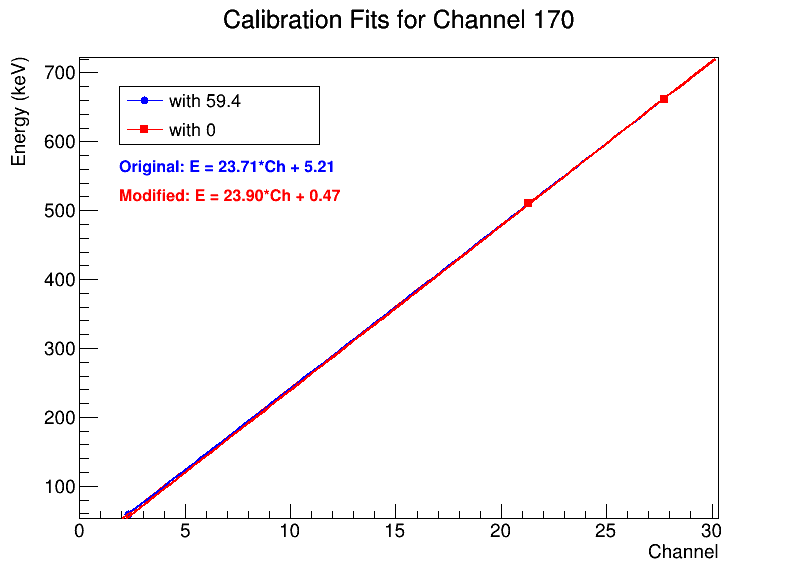}
        \put(80,10){\textbf{(b)}}
    \end{overpic}
    
    \caption{Three-point calibration fit examples for a single detector channel, corresponding to: (a) front crystal, and (b) back crystal.}
    \label{fig:channel_170_comparison}
\end{figure}

To determine whether the exclusion of the 59.4~keV point significantly changes the calibration accuracy, two independent calibration fits were generated for channels where all three peaks were available: (i) using \{0, 511, 662\}~keV and (ii) using \{59.4, 511, 662\}~keV. As shown in Figures~\ref{fig:channel_170_comparison}, the difference between the two fits was negligible and far below the intrinsic energy resolution of the detector. Consequently, the calibration based on \{0, 511, 662\}~keV was adopted for all channels in both layers.

\begin{figure}[!ht]
    \centering
    
    \begin{overpic}[width=0.8\linewidth]{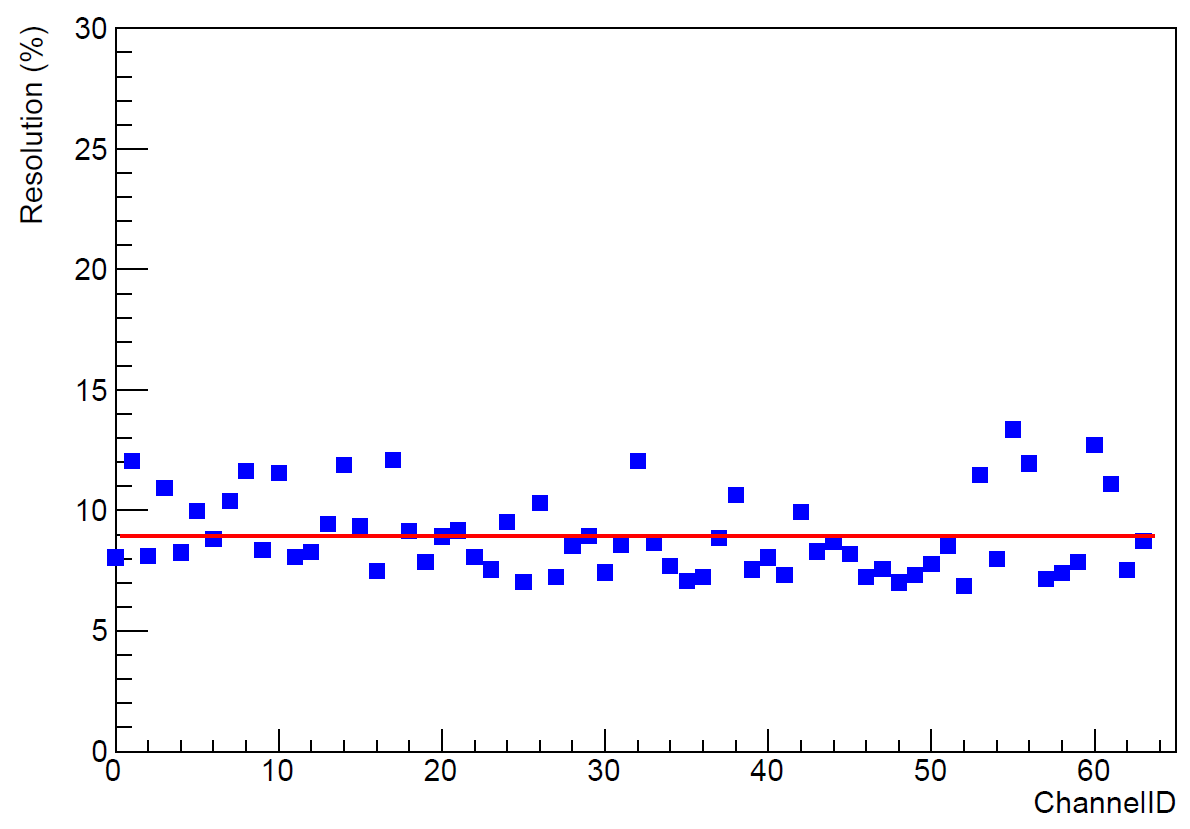}
        \put(90,10){\textbf{(a)}}
    \end{overpic}
    
    \vspace{0.5em}
    
    \begin{overpic}[width=0.8\linewidth]{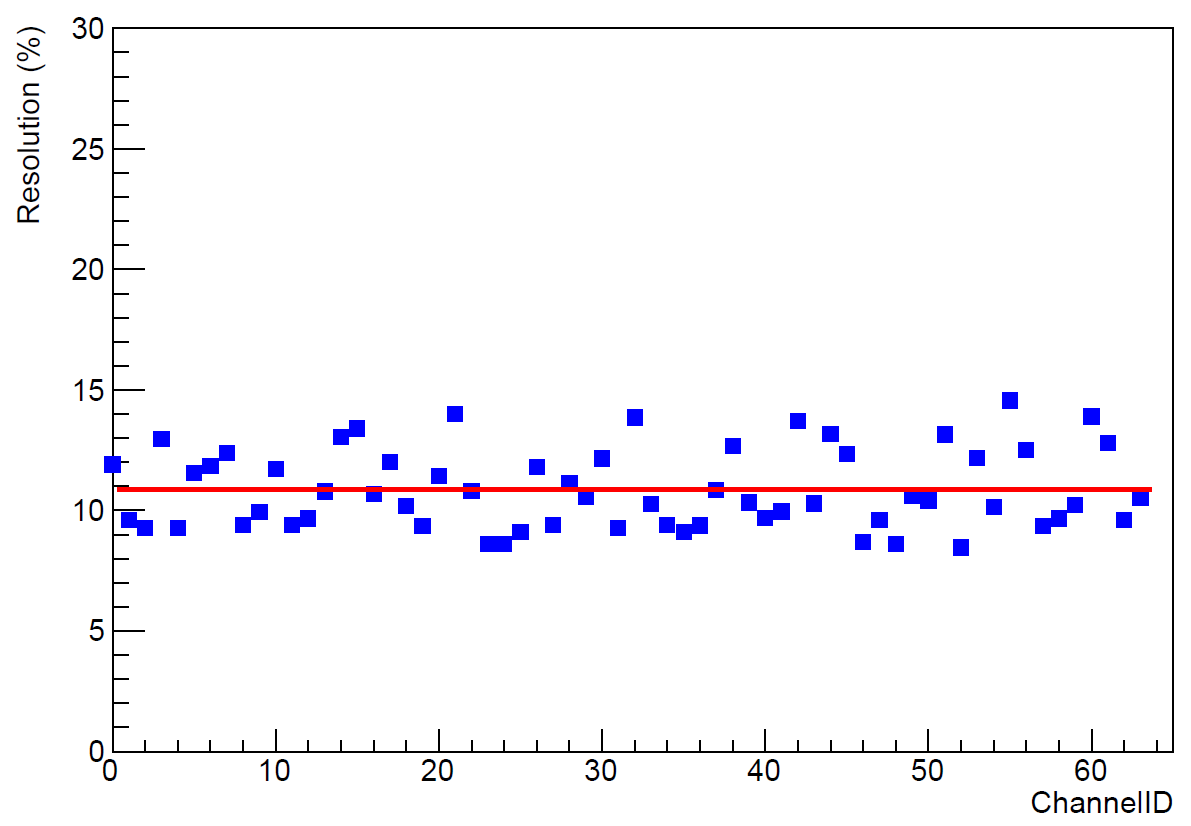}
        \put(90,10){\textbf{(b)}}
    \end{overpic}
    
   \caption{Energy resolution of individual pixels when the (a) front layer, and (b) back layer is irradiated with a collimated beam of \textsuperscript{137}Cs gammas.}
\label{fig:Energy_resolution}
\end{figure}

Gaussian fits to the photopeaks at 662 keV, yielded average energy resolutions of $(8.9 \pm 1.9)\%$ for the front layer and $(10.8 \pm 1.6)\%$ for the back layer, respectively as shown in the Fig.~\ref{fig:Energy_resolution}, where the quoted uncertainties represent one standard deviation. 
In addition to the superior energy resolution, we observed that the front pixels yield on average 20\% higher signal, than the back pixels of the same elements. From this, we conclude that the difference must come from the number of optical photons reaching the SiPM active area. In the case of the front pixel, the scintillation photons are reflected at five sides, and transmitted backward. In case of the back pixel, the scintillation photons are reflected at four sides, a partially transmitted forward and partially towards the SiPM. This presumably results in lower light yield in the SiPM, and consequently lower energy resolution for the crystals in the back-layer.

Energy calibration was determined by the coefficients extracted from the dedicated measurements, as described above. For the imaging measurements, minor shifts in reconstructed photo-peak positions were observed, likely due to temperature variations. These shifts were corrected using a common energy correction factor, aligning all peak positions closely with the expected peak position.

\subsection{Data Selection and Analysis} \label{sec:Data_selection_and_analysis}

Accurate reconstruction of Compton events requires stringent event selection to ensure that only physically valid interactions contribute to image formation. The analysis was performed in four sequential steps: preliminary energy selection, transaxial distance verification, interaction sequence determination, and final energy window selection.

Since the calibration study demonstrated that the front detector layer exhibits a higher energy response, all events were initially calibrated assuming that the interaction occurred in the front layer. A lower energy threshold of 120~keV was then applied to every channel to suppress electronic noise. In addition, only events in which exactly two SiPM pixels were triggered simultaneously were retained as Compton event candidates.

\begin{figure}
    \centering
    
    \begin{overpic}[width=0.28\linewidth]{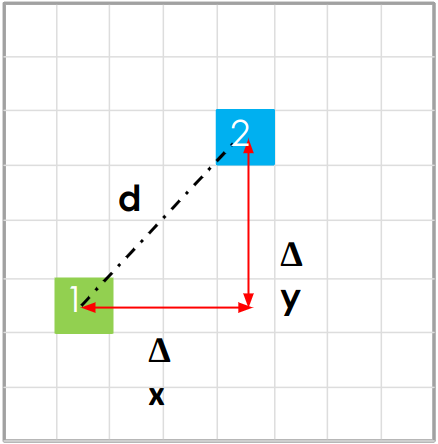}
        \put(80,10){\textbf{(a)}}
    \end{overpic}
    \hspace{0.02\linewidth}
    \begin{overpic}[width=0.65\linewidth]{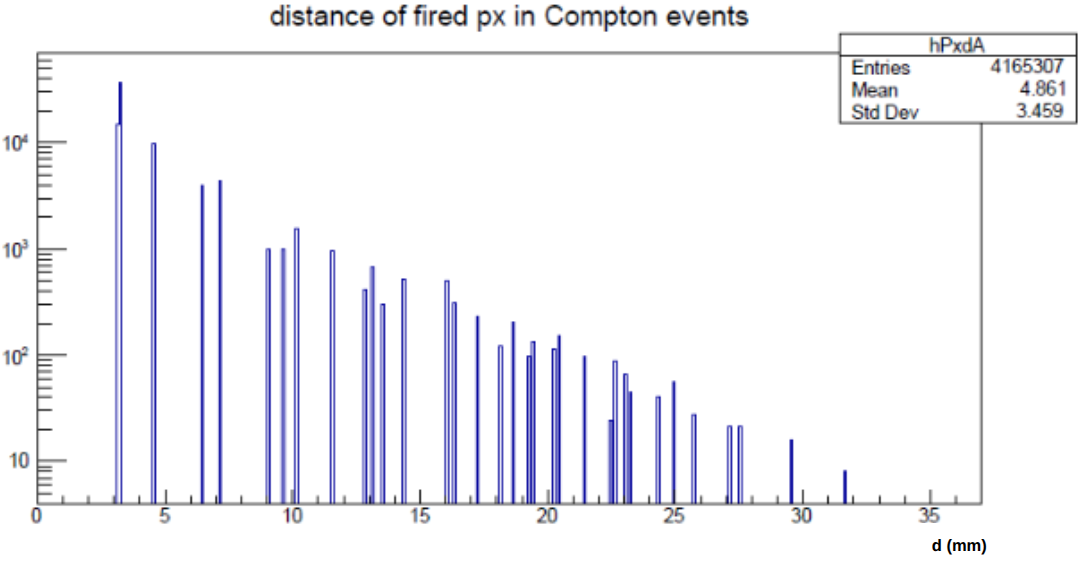}
        \put(80,10){\textbf{(b)}}
    \end{overpic}
    
    \caption{(a) Schematic representation of the transaxial distance definition, (b) number of reconstructed Compton events as a function of transaxial distance of the fired pixels.}
    \label{fig:reconstructed_events_vs_distance}
\end{figure}

The second selection criterion was based on the transaxial distance between the two fired pixels. This requirement served two purposes. First, it suppressed events affected by optical crosstalk between neighboring scintillator elements. Second, it enforced the geometrical constraints imposed by Compton kinematics together with the 120~keV lower energy threshold. Figure~\ref{fig:reconstructed_events_vs_distance}(a) illustrates the definition of the transaxial distance, while Fig.~\ref{fig:reconstructed_events_vs_distance}(b) presents the number of reconstructed events as a function of this distance for a \textsuperscript{137}Cs source. Based on these distributions, minimum transaxial distances of $d_{\mathrm{min}}=10$~mm for \textsuperscript{137}Cs and $d_{\mathrm{min}}=13$~mm for \textsuperscript{22}Na were selected.


\begin{figure}
    \centering
    \includegraphics[width=0.95\linewidth]{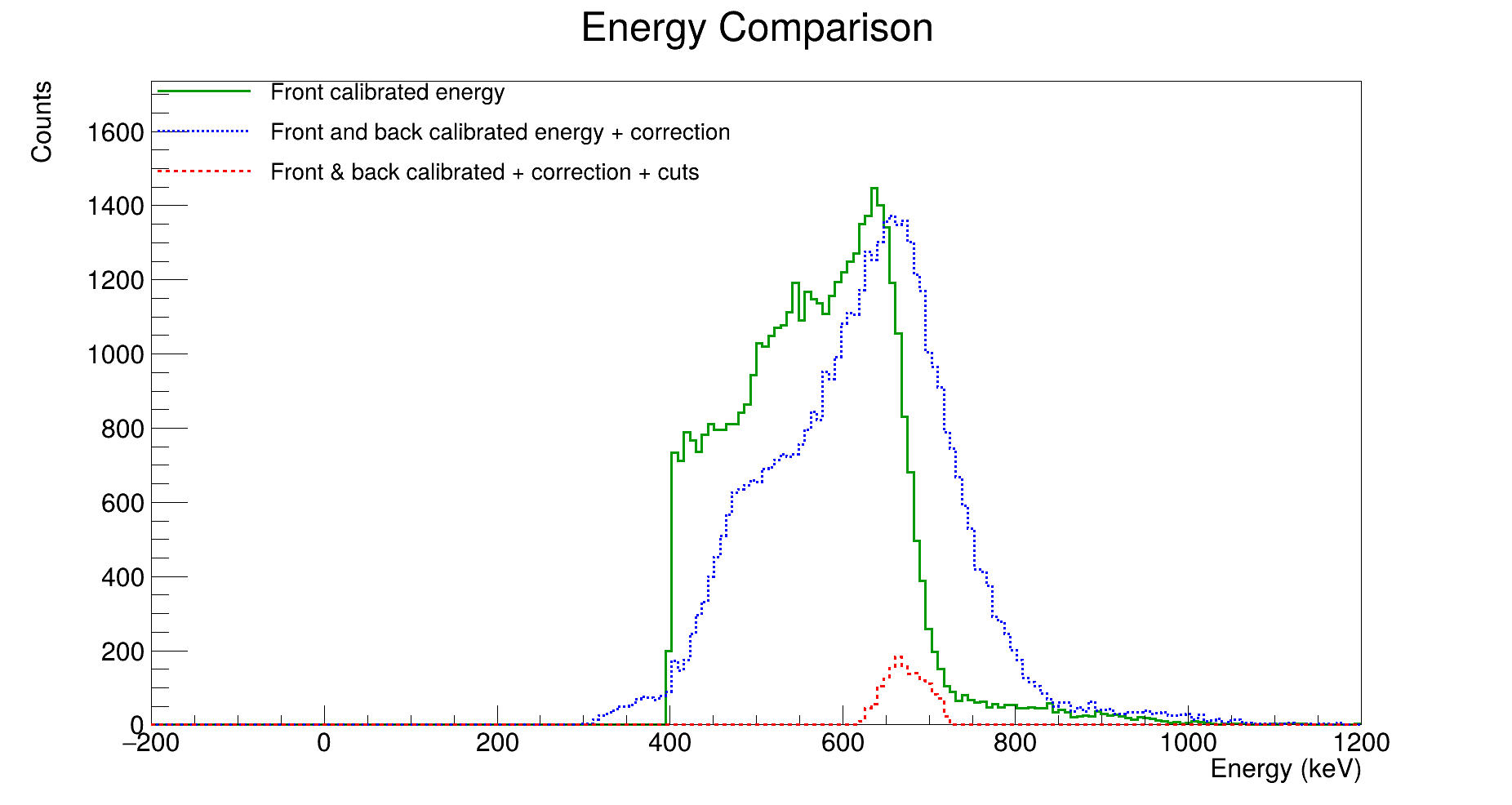}
    \caption{Reconstructed sum of the response of the two fired SiPM pixels for 662~keV gamma photons, at different analysis steps (see text).}
    \label{fig:sipm_two_pixel_sum_662}
\end{figure}

\begin{figure}
    \centering
    \includegraphics[width=0.95\linewidth]{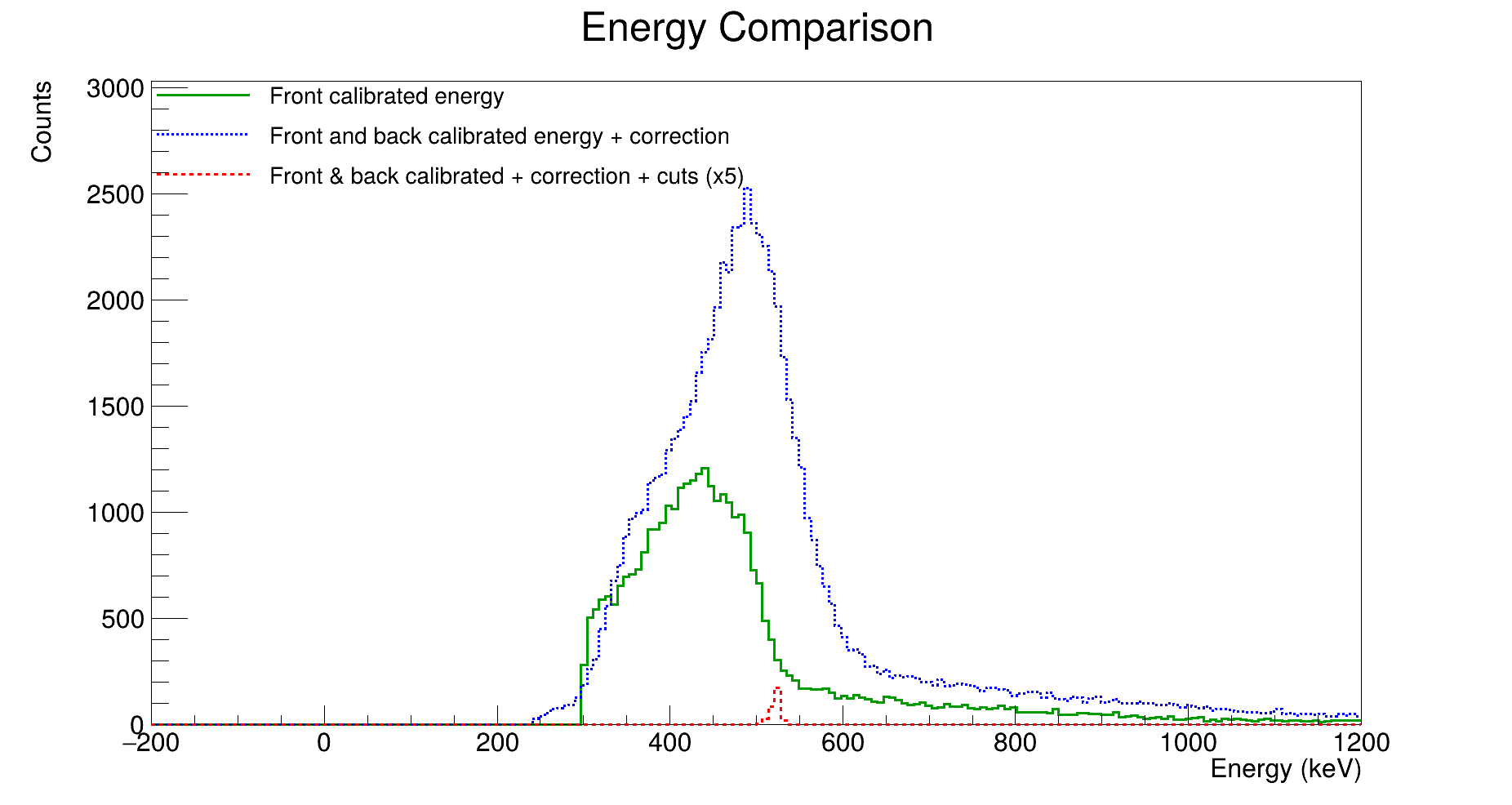}
    \caption{Reconstructed sum of the response of the two fired SiPM pixels for 511~keV gamma photons, at different analysis steps (see text).}
    \label{fig:sipm_two_pixel_sum_511}
\end{figure}

Because all events were initially calibrated using the front-layer calibration curve, interactions occurring in the back layer had an underestimated energy due to the lower light response. Consequently, the summed energy of the two fired pixels was approximately $10\%$ lower than the incident gamma-ray energy. This effect is visible in the green spectra of Figs.~\ref{fig:sipm_two_pixel_sum_662} and~\ref{fig:sipm_two_pixel_sum_511} for the 662~keV and 511~keV sources, respectively. 

To correctly identify the interaction sequence, the analysis was restricted to the forward-scattering region. According to Compton scattering kinematic for the gamma energies of interest, 
a forward-scattered photon deposits less energy during its first interaction in the scatterer, leaving a higher fraction of energy to the scattered photon. Therefore, for events with two simultaneously triggered pixels, the pixel with the lower deposited energy was assigned to the first interaction in the scatterer, while the pixel with the higher deposited energy was assigned to the second interaction in the absorber. This criterion enabled an unambiguous determination of the interaction sequence.



Once the interaction order had been established, the appropriate calibration curve was applied to the back-layer pixel. The corrected energies of both interactions were then summed to obtain the reconstructed gamma-ray energy. The resulting spectra are shown by the blue curves in Figs.~\ref{fig:sipm_two_pixel_sum_662} and~\ref{fig:sipm_two_pixel_sum_511} for \textsuperscript{137}Cs and \textsuperscript{22}Na, respectively.

Finally, an asymmetric energy window extending from $-\sigma$ below the full-energy peak to $+2\sigma$ above, was selected. This selection suppressed events with incomplete energy deposition caused by interactions within the light guide while simultaneously rejecting most incorrectly ordered backscattering events, as confirmed by simulation (see Section~\ref{sec:simulation}). The accepted events after this final selection are represented by the red spectra in Figs.~\ref{fig:sipm_two_pixel_sum_662} and~\ref{fig:sipm_two_pixel_sum_511}.

\begin{figure}[!ht] 
    \centering
    
   \begin{overpic}[width=0.995\linewidth,height=5cm]{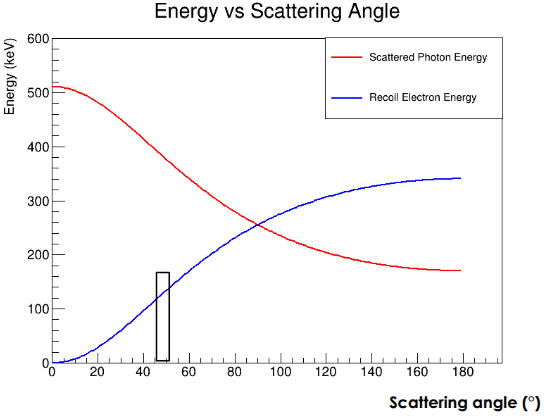}
    \put(85,12){\textbf{(a)}}
\end{overpic}
    
    \vspace{0.5em}
    
    \begin{overpic}[width=0.995\linewidth]{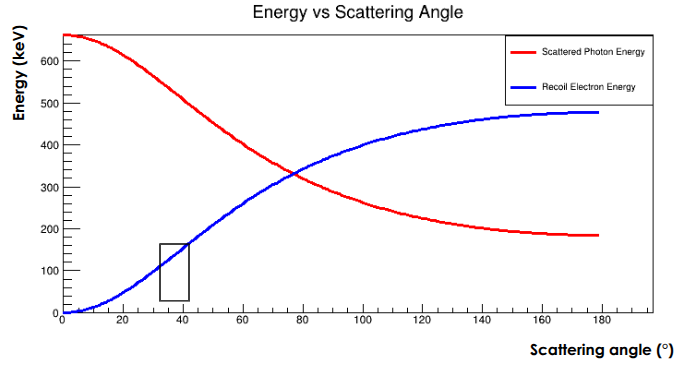}
        \put(85,10){\textbf{(b)}}
    \end{overpic}
    
    \caption{Energy distribution of the recoil electron and scattered photon as a function of scattering angle: (a) for a 511 keV gamma photon, (b) for a 662 keV gamma photon. The region inside the black box is the energy range of the recoil electron energy used for event reconstruction in the analysis for the experiment.
    }
    \label{fig:compton_energy_distribution_662}
\end{figure}

Finally, the permissible energy ranges for the scatterer and absorber were optimized separately for each gamma-ray energy. Figure~\ref{fig:compton_energy_distribution_662}(a) shows the energy-sharing distribution for 511~keV photons from \textsuperscript{22}Na, while Fig.~\ref{fig:compton_energy_distribution_662}(b) presents the corresponding distribution for 662~keV photons from \textsuperscript{137}Cs. These optimized selection regions maximize the number of correctly ordered Compton events while minimizing contamination from backscattered events.


\begin{figure}
    \centering
    \includegraphics[width=0.95\linewidth]{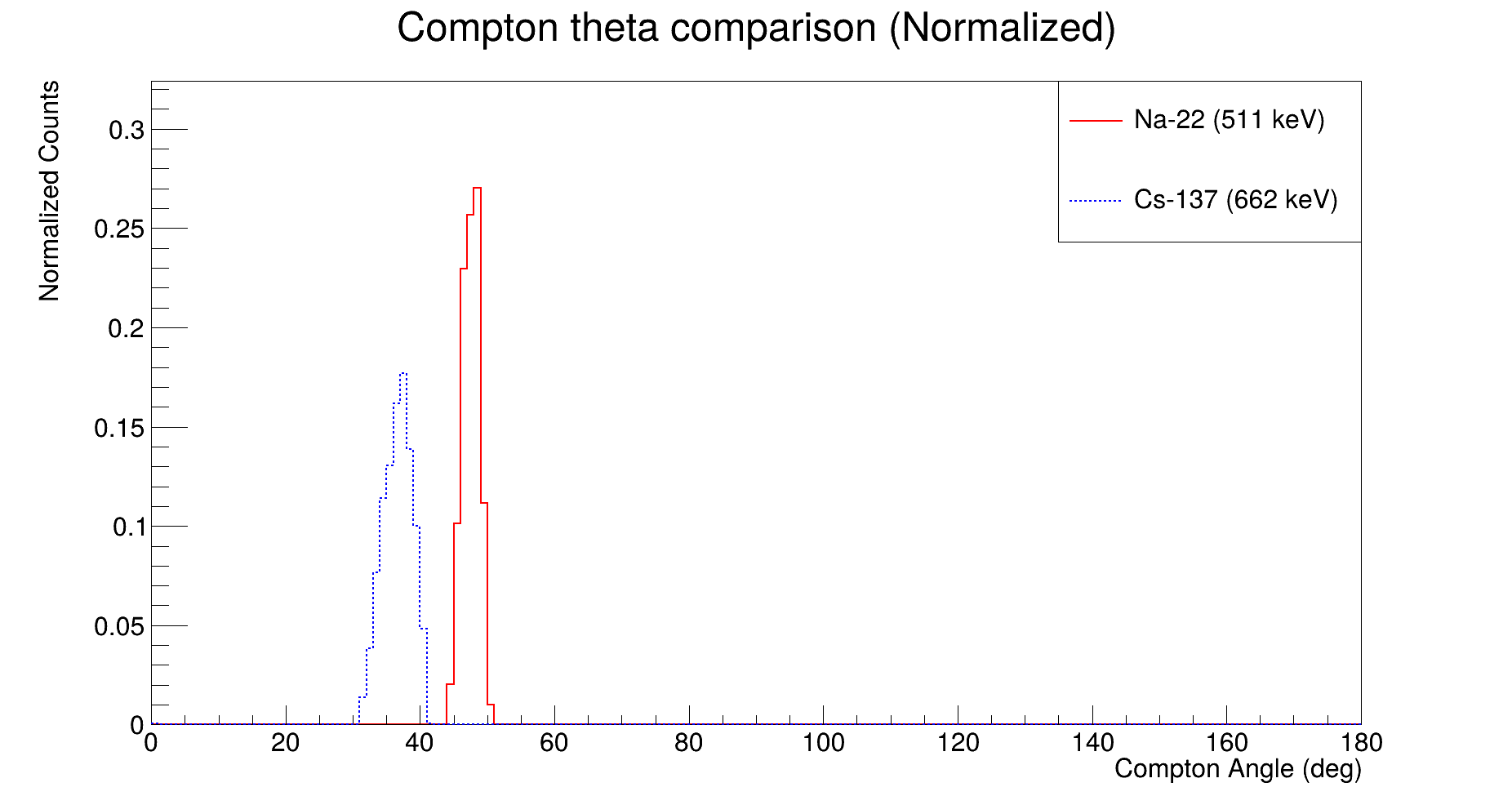}
    \caption{Distribution of Compton scattering angles for all reconstructed events for \textsuperscript{22}Na and \textsuperscript{137}Cs placed centrally at 10 cm from the detector.}
    \label{fig:compton_angle_distribution_from_exp}
\end{figure}

After applying all selection criteria, the reconstructed Compton scattering angle distributions for the 511~keV and 662~keV gamma-ray sources are presented in Fig.~\ref{fig:compton_angle_distribution_from_exp}.


\subsection{Image Reconstruction}
\label{sec:image_reconstruction}

Reconstruction of the incident gamma-ray direction in a Compton camera requires accurate measurements of the interaction positions and deposited energies from at least two detector layers. Each valid Compton event constrains the possible source locations to a conical surface, known as the Compton cone. By accumulating information from multiple events, the spatial distribution of the radioactive source can be reconstructed. The achievable image quality depends on detector resolution, event statistics, and the reconstruction algorithm employed.

In this work, an image reconstruction alghorithm based on list-mode MLEM (LM-MLEM) was applied. For each reconstructed and selected Compton events, X, Y, and Z coordinates of the interacted pixels, along with the deposited energies, were recorded in list-mode format. This dataset provided the input for reconstruction algorithm.

The MLEM algorithm is an iterative reconstruction method that improves image quality by incorporating a statistical model of the detection process. In this framework, each voxel activity, $f_j$, represents the expected photon emission from voxel $j$, while the measured data, $p_i$, are modeled as Poisson-distributed random variables,

\begin{equation}
P(p_i|f)=\frac{\bar{p}_i^{\,p_i}e^{-\bar{p}_i}}{p_i!},
\end{equation}

where

\begin{equation}
\bar{p}_i=\sum_j H_{ij}f_j
\end{equation}

is the expected number of detected events for measurement $i$, and $H_{ij}$ is the system matrix describing the probability that a photon emitted from voxel $j$ contributes to measurement $i$.

Following the Expectation Maximization formalism introduced by Shepp and Vardi~\cite{shepp1982maximum}, the voxel values are iteratively updated according to

\begin{equation}
f_j^{k+1}
=
\frac{f_j^k}{\sum_i H_{ij}}
\sum_i
\frac{p_iH_{ij}}
{\sum_l H_{il}f_l^k}.
\end{equation}

For list-mode Compton camera data, a more suitable formulation is given by Wilderman \textit{et al.}~\cite{wilderman2001improved},

\begin{equation}
f_j^{k+1}
=
\frac{f_j^k}{s_j}
\sum_{i(\cap j)}
\frac{t_{ij}}
{\sum_{l(\cap i)} t_{il}f_l^k},
\end{equation}

where $t_{ij}$ represents the probability that event $i$ originates from voxel $j$, and

\begin{equation}
s_j=\sum_i t_{ij}
\end{equation}
denotes the sensitivity of voxel $j$.

In this work, the reconstruction is performed over a \(200 \times 200\) voxel grid with a spatial pitch of \(1\,\mathrm{mm}\), defining a field of view of \(200 \times 200\,\mathrm{mm}^2\). A total of 20 MLEM iterations are applied. To reduce computational complexity, a simplified geometric system matrix is used, where \(t_{ij} = 1\) for voxels intersecting the corresponding Compton cone and \(t_{ij} = 0\) otherwise. This binary approximation neglects detailed physical effects such as attenuation and detector response, but captures the dominant geometric contribution of each event.

\section{Monte Carlo Simulation of the Detector}
\label{sec:simulation}

Monte Carlo simulations were performed to optimize the event selection and image reconstruction of the assembled single-plane readout CC, to evaluate it intrinsic efficiency, and to provide a reference for comparison with experimental results. The simulations were carried out using the \textsc{Geant4} toolkit~\cite{geant4}, which offers a detailed and well-validated framework for modeling particle interactions with matter, particularly for gamma-ray transport and electromagnetic processes.

\subsection{Detector Geometry and Materials}

The detector geometry was modeled to closely replicate the experimental prototype. An $8 \times 8$ pixelated array was implemented, where each pixel consisted of two GAGG:Ce scintillator crystals with dimensions of $3~\mathrm{mm} \times 3~\mathrm{mm} \times 3~\mathrm{mm}$, functioning as the scatterer and absorber. The two crystals were coupled via a plexiglass light guide with a length of $20~\mathrm{mm}$. 
One side of the detector matrix was coupled to a single silicon piece with dimensions $25.8 \times 25.8 \times 1.4\ \mathrm{mm}^3$, representing the SiPM array used in the experimental setup. Optical photon transport was not explicitly simulated; instead, idealized energy deposition and interaction positions were recorded to assess the intrinsic performance of the setup.


\subsection{Physics List and Event Generation}

Electromagnetic interactions were modeled using the standard \textsc{Geant4} electromagnetic physics list, which includes Compton scattering, photoelectric absorption, and Rayleigh scattering. Gamma-ray sources emitting monoenergetic photons at 511~keV and 662~keV were simulated using a single-particle gun. The 511~keV photons represent annihilation radiation from \textsuperscript{22}Na, while the 662~keV photons correspond to emissions from \textsuperscript{137}Cs.
Photons were generated isotropically from point-like sources positioned at various locations within the detector field of view. For each configuration, a sufficient number of primary events were simulated to obtain stable estimates of the detector efficiency, energy deposition, and image reconstruction performance.

\subsection{Optimization of Event Selection}
\label{subsec:event_selection}
To evaluate the performance of the assembled single-plane readout CC under realistic conditions, the experimental constraints were closely reproduced. In particular the detector energy response was smeared according to the experimental resolution, and the reconstructed pixel positions were obtained by taking their mid-points. 

To emulate the single-sided readout scheme, the energy deposits from the front and back scintillator of within each detector element were first summed according to: 
\begin{equation}
E_\text{tot} = E_\text{Layer1} + E_\text{Layer2}.
\end{equation}
The $E_\text{tot}$ were then smeared using a Gaussian, with a referent energy resolution of 12\% at 662 keV, assuming the $\sqrt{E}$ scaling:
\begin{equation}
\sigma(E_\text{tot}) = \frac{0.12}{2.35} \sqrt{662\ \text{keV} \cdot E_\text{tot}}.
\end{equation}
The final simulated energy was then sampled from the Gaussian distribution:
\begin{equation}
E_\text{smeared} \sim \mathcal{N}(E_\text{tot}, \sigma(E_\text{tot})).
\end{equation}


Event selection was limited by the experimental low-energy threshold of 120~keV, imposed to suppress electronic noise. Upper bounds on the energy deposited in the low-energy pixel were defined using Compton scattering kinematics to minimize contributions from backscattering. In addition, a minimum inter-pixel distance cut was applied to reject closely spaced interactions that degrade angular resolution. A distance cut of 13~mm for $^{22}$Na and 10~mm for $^{137}$Cs, consistent with the experimental analysis, was adopted as the baseline configuration.

Reconstructed events were classified by tracking event identifiers and comparing them with the ideal simulation case, in which full interaction information was available. Three dominant categories were identified: correctly reconstructed forward-scattering events, backward-scattering events, and events involving interactions in the light guide. The latter two categories may be misidentified and therefore represent a background contributing to image degradation.

To suppress this background as effectively as possible, the energy spectra of each contribution, along with the two-dimensional energy correlations, were analyzed using interaction-level information obtained from the simulation. This study revealed that 
interactions in the light guide produce a systematic shift toward lower total deposited energy. In the forward scattering region (black boxes in Figure  \ref{fig:compton_energy_distribution_662}), misidentified backward-scattering events will exhibit a shift to lower total energy deposition, due to Compton kinematics.

Based on this observation, the selection criteria were refined by introducing an asymmetric cut on the total deposited energy, defined as $1\sigma$ below the peak and $2\sigma$ above it or an absorber energy threshold was applied, requiring the deposited energy in the absorber to be greater than 380~keV for $^{22}$Na and greater than 500~keV for $^{137}$Cs. The final selection criteria applied are listed in Table \ref{tab:Combined_analysis_parameters}. These optimized cuts increased the fraction of correctly reconstructed events to about 81\% while suppressing background contributions, as listed in  Table~\ref{tab:event_classification_after_cuts}. 


\begin{table}
\caption{Energy selection and detector parameters for $^{137}$Cs and $^{22}$Na data analysis}
\label{tab:Combined_analysis_parameters}
\centering
\begin{tabular}{l c c}
\hline
\textbf{Parameter} & \textbf{$^{137}$Cs} & \textbf{$^{22}$Na} \\
\hline
Inter-pixel distance       & $> 10~\mathrm{mm}$ &  $> 13~\mathrm{mm}$\\
Low energy range (scatterer) & 120--160~keV & 120--140~keV \\
High energy range (absorber) & $> 500~\mathrm{keV}$ & $> 380~\mathrm{keV}$ \\
Sum energy range           & 642--700~keV & 494--546~keV \\
\hline
\end{tabular}
\end{table}

\begin{table}
\centering
\caption{Classification of reconstructed events after applying optimized energy cuts for 511~keV and 662~keV gamma rays.}
\label{tab:event_classification_after_cuts}
\begin{tabular}{lcc}
\hline
\textbf{Event Type} & \textbf{511 keV} & \textbf{662 keV} \\
\hline
Correctly reconstructed  & 80\% & 81\% \\
Back-scatter events & $11\%$ & $8\%$ \\
Light guide interactions & 9\% & $11\%$ \\
\hline
\end{tabular}
\end{table}



\section{Results}

\subsection{Imaging of the $^{137}$Cs point-source}



\begin{figure}
    \centering
    \includegraphics[width=0.95\textwidth]{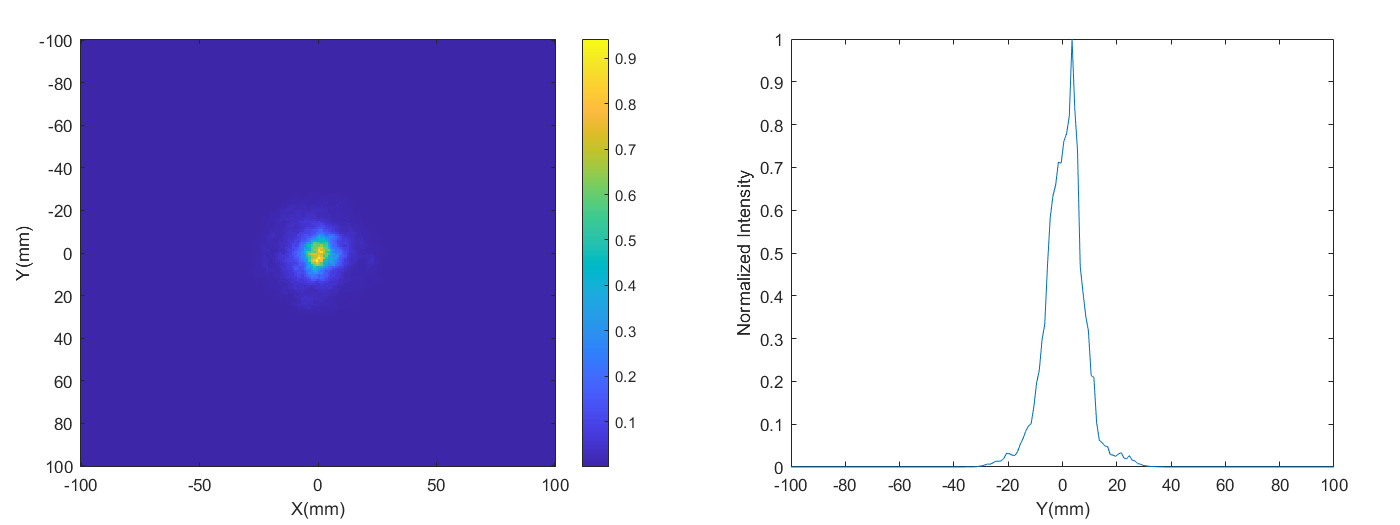}
    \caption[The MLEM image and its X and Y-profile of $^{137}$Cs source placed at 100 mm away from the face of the detector at the centre]{(Left) MLEM reconstructed image and (right) Y-profile for a $^{137}$Cs source positioned at $(x,y)=(0,0)$ mm, 100 mm from the detector face.}
    \label{fig:Cs137_exp_centre_MLEM}
\end{figure}

\begin{figure}
    \centering
    \includegraphics[width=0.95\textwidth]{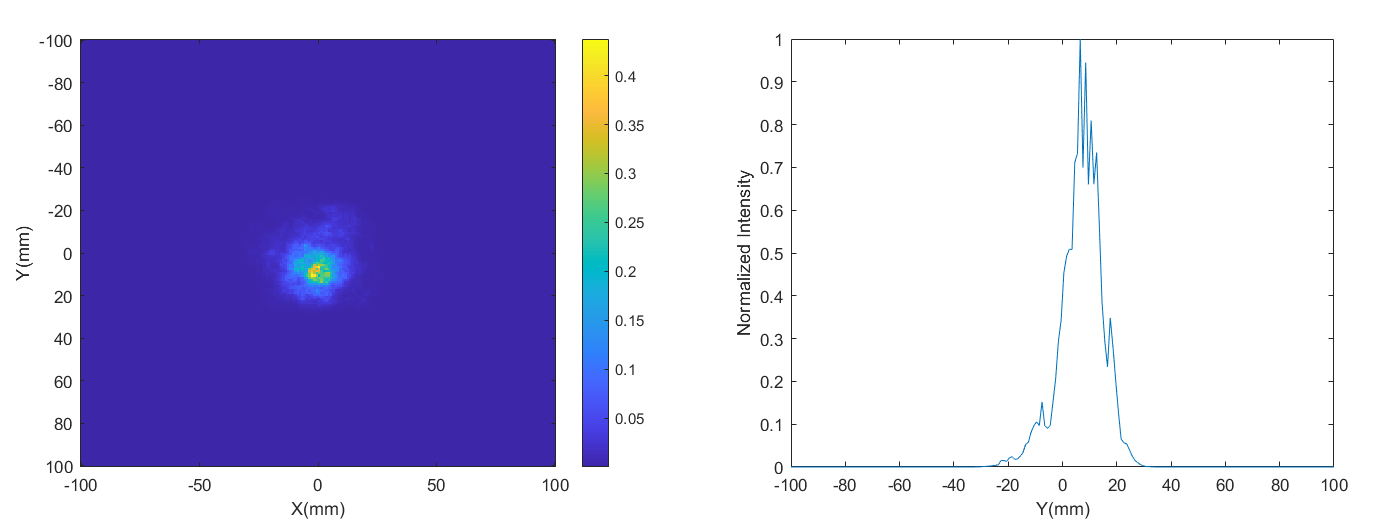}
    \caption[The MLEM image and its X and Y-profile of $^{137}$Cs source placed at 100 mm away from the face of the detector at (x = 0 mm, y = 10 mm)]{(Left) MLEM reconstructed image and (right) Y-profile for a $^{137}$Cs source located at $(x,y)=(0,10)$ mm and 100 mm from the detector face.}
    \label{fig:Cs137_exp_1cmright_MLEM_profile}
\end{figure}

\begin{figure}
    \centering
    \includegraphics[width=0.95\textwidth]{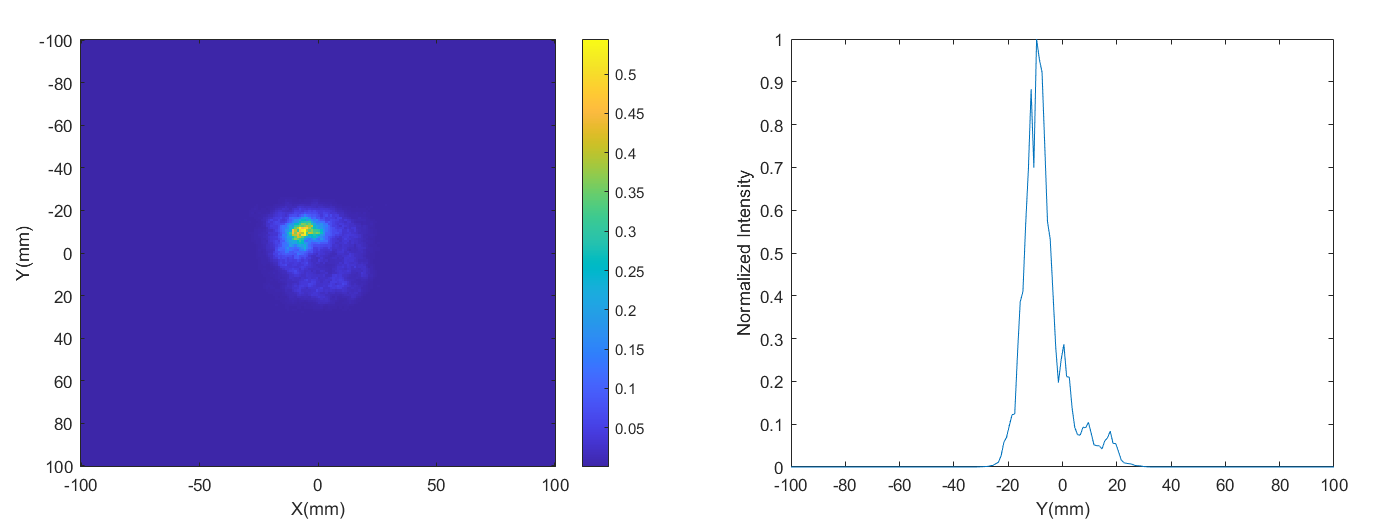}
    \caption[The MLEM image and its X and Y-profile of $^{137}$Cs source placed at 100 mm away from the face of the detector at (x = -8 mm, y = -8 mm)]{(Left) MLEM reconstructed image and (right) Y-profile for a $^{137}$Cs source positioned at $(x,y)=(-8,-8)$ mm, 100 mm from the detector face.}
    \label{fig:Cs137_exp_8*8mm_topleft_MLEM}
\end{figure}

The origin (x,y,z)=(0,0,0) is defined at the centre of the detector's front surface, with the x- and y-axes lying in the detector plane and the z-axis normal to the detector surface and pointing toward the source region.
With the source positioned as described in Table \ref{tab:measurement_summary}, and following the selection criteria listed in Table~\ref{tab:Combined_analysis_parameters}, the reconstructed images are shown in Figures \ref{fig:Cs137_exp_centre_MLEM}, \ref{fig:Cs137_exp_1cmright_MLEM_profile}, and \ref{fig:Cs137_exp_8*8mm_topleft_MLEM}.
Images confirm that the source position is correctly reconstructed in both X and Y directions with the positioning and imaging accuracy.

\begin{figure}
    \centering
    \includegraphics[width=0.5\textwidth]{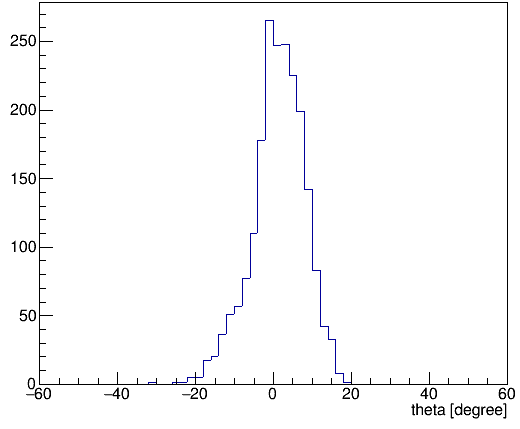}
    \caption{ARM distribution for a $^{137}$Cs point-source positioned 100 mm from the detector at the central axis.}
    \label{fig:Cs137_exp_centre_ARM}
\end{figure}

The Angular Resolution Measure (ARM) for a centrally placed source at 100 mm was $14.3^\circ$, as shown for example in Figure~\ref{fig:Cs137_exp_centre_ARM}. Off-axis measurements, including shifts of 10 mm along $Y$ and top-left positions (-8 mm, -8 mm), maintained similar ARM and efficiency values, reported in Table~\ref{tab:reco_exp_results}. 

\subsection{Imaging of the $^{22}$Na point-source}

\begin{figure}
    \centering
    \includegraphics[width=0.95\textwidth]{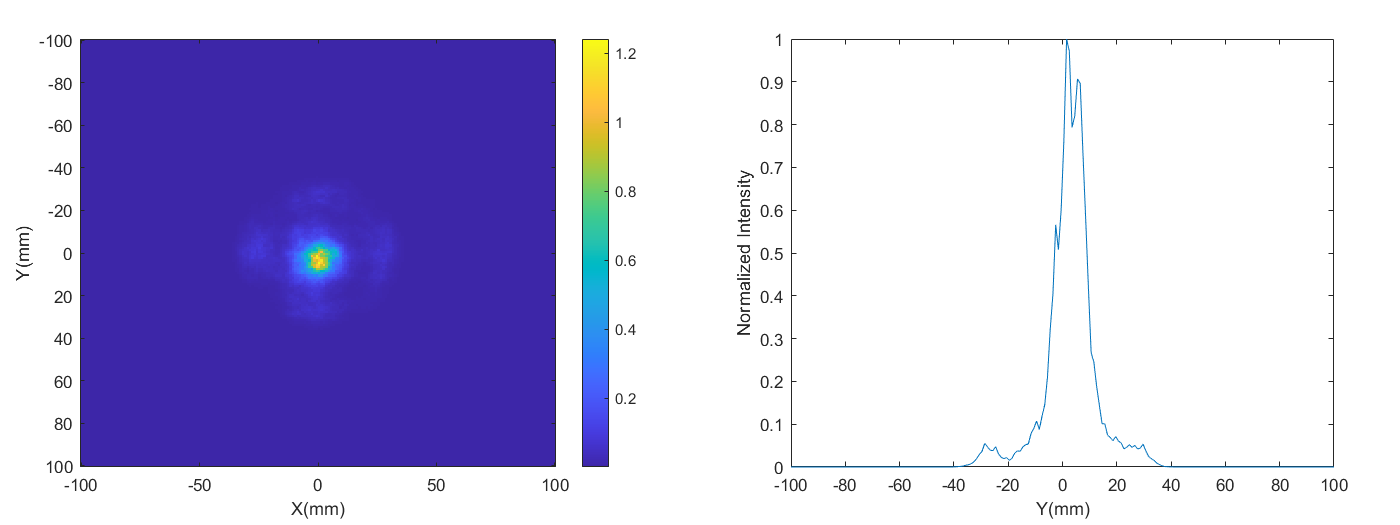}
    \caption[The MLEM image  and its X and Y-profile of $^{22}$Na source placed at 100 mm away from the face of the detector at the centre]{(Left) MLEM reconstructed image and (right) Y-profile for a $^{22}$Na source positioned at $(x,y)=(0,0)$ mm, 100 mm from the detector face.}
    \label{fig:Na22_exp_centre_MLEM}
\end{figure}

\begin{figure}
    \centering
    \includegraphics[width=0.95\textwidth]{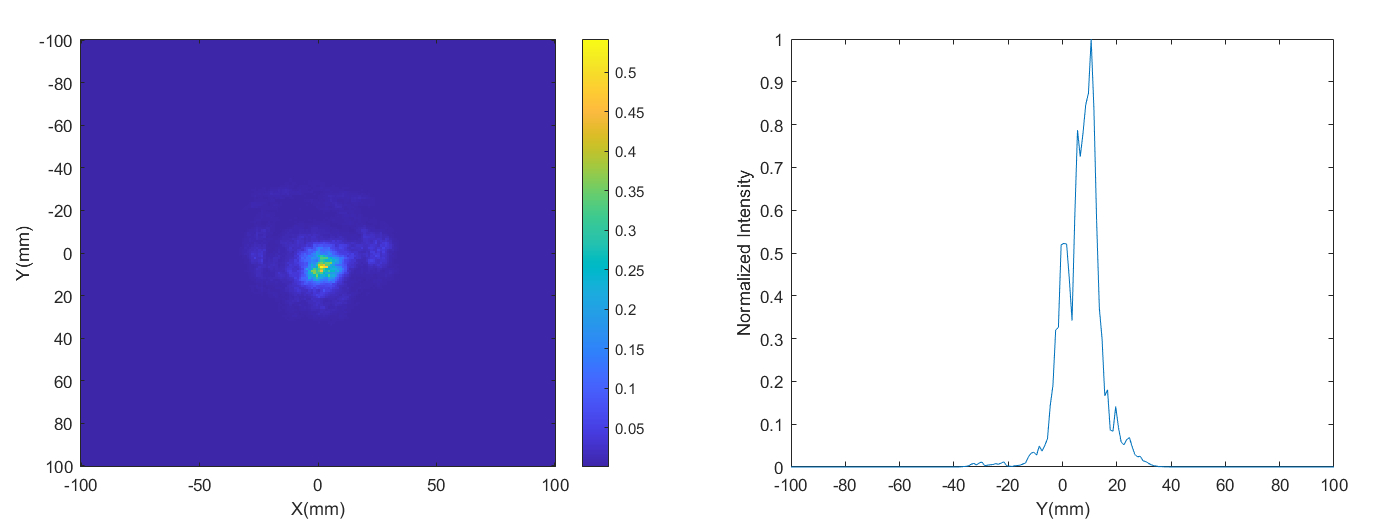}
    \caption[The MLEM image  and its X and Y-profile of $^{22}$Na source placed at 100 mm away from the face of the detector at Y = 10 mm]{(Left) MLEM reconstructed image and (right) Y-profile for a $^{22}$Na source positioned at $(x,y)=(0,10)$ mm, 100 mm from the detector face.}
    \label{fig:Na22_exp_y1_MLEM}
\end{figure}

\begin{figure}
    \centering
    \includegraphics[width=0.95\textwidth]{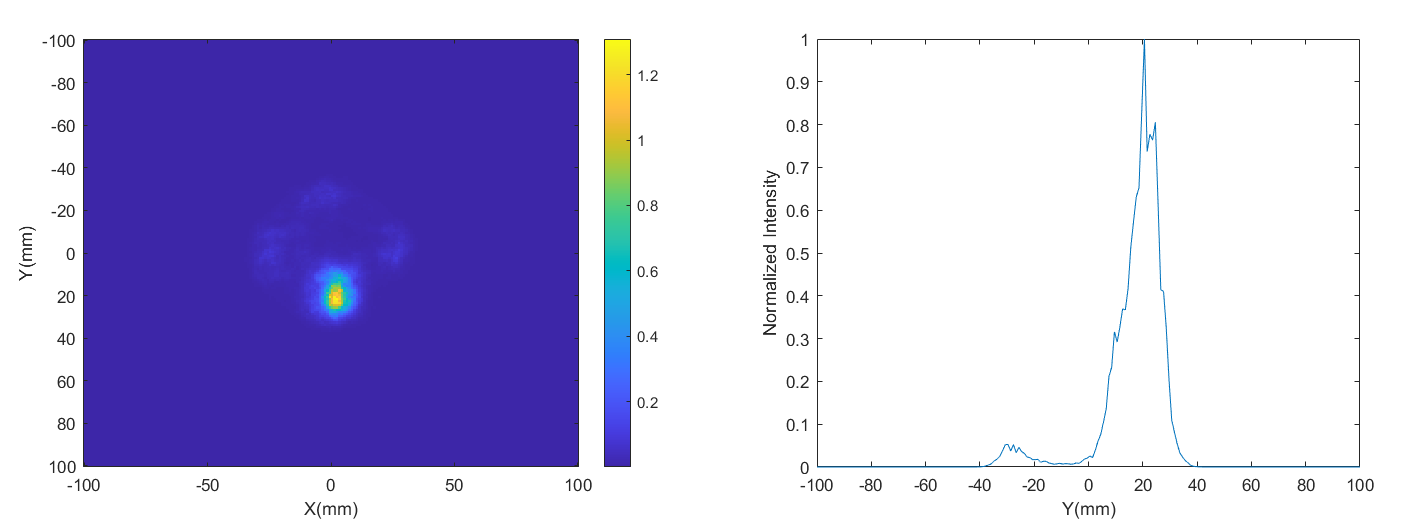}
    \caption[The MLEM image  and its X and Y-profile of $^{22}$Na source placed at 100 mm away from the face of the detector at Y = 30 mm]{(Left) MLEM reconstructed image and (right) Y-profile for a $^{22}$Na source positioned at $(x,y)=(0,30)$ mm, 100 mm from the detector face.}
    \label{fig:Na22_exp_y3_MLEM}
\end{figure}

The $^{22}$Na point-source was positioned at the center and at 10~mm and 30~mm along the Y-axis, 100~mm in front of the detector. Reconstructed respective images are shown in Figures~\ref{fig:Na22_exp_centre_MLEM}--\ref{fig:Na22_exp_y3_MLEM}. They show that the source position is correctly reconstructed (within the given precision) for the central position and for the source at Y=10 mm off axis. For the last case with Y=30 mm off axis, the reconstructed off-axis shift is underestimated by $\sim$25\%. This results from the specific event selection conditions as discussed in Section \ref{discussion}. 

The ARM values, as well as the intrinsic efficiencies for each case are summarized in the Table \ref{tab:reco_exp_results}, and compared to the values derived from the corresponding simulations. 

\begin{figure}
    \centering
    \includegraphics[width=0.5\textwidth]{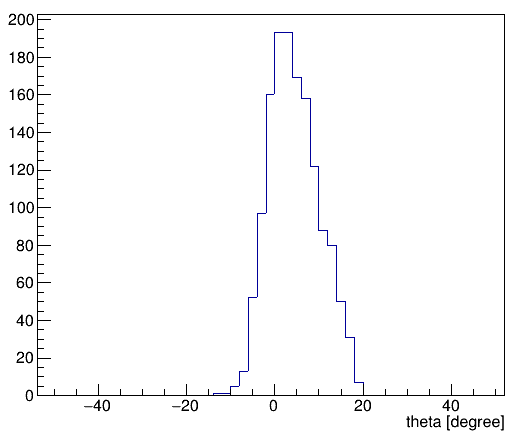}
    \caption[The ARM plot of the $^{22}$Na source placed at 100 mm away from the face of the detector at the centre]{The ARM of $^{22}$Na source placed at 100 mm away from the face of the detector at the centre.}
    \label{fig:ARM_Na_centre_exp}
\end{figure}

\begin{table}
\caption{Detector characteristics for imaging of $^{22}$Na and $^{137}$Cs sources}
\label{tab:reco_exp_results}
\centering
\renewcommand{\arraystretch}{1.2}
\resizebox{\linewidth}{!}{%
\begin{tabular}{l c c c c c c c}
\hline
\textbf{Source} & 
\textbf{True (x,y)} & 
\textbf{Reco (x,y)} & 
\textbf{$\sigma_x$} & 
\textbf{$\sigma_y$} & 
\textbf{ARM (exp.)} & 
\textbf{ARM (sim.)} & 
\textbf{Intrinsic} \\

 & 
\textbf{[mm]} & 
\textbf{[mm]} & 
\textbf{[mm]} & 
\textbf{[mm]} & 
\textbf{(FWHM)} & 
\textbf{(FWHM)} & 
\textbf{efficiency} \\
\hline

$^{137}$Cs & 0, 0       & $-0.2,\,1.7$   & 4.3 & 5.2 & $14.3 \pm 0.5^\circ$ & $15.9 \pm 0.4^\circ$ & $(4.3 \pm 0.3)\times10^{-5}$ \\
$^{137}$Cs & 0, 10      & $-0.7,\,7.4$   & 5.4 & 5.4 & $15.0 \pm 0.6^\circ$ & $13.3 \pm 0.4^\circ$ & $(4.6 \pm 0.3)\times10^{-5}$ \\
$^{137}$Cs & $-8, -8$   & $-6.9,\,-9.5$  & 4.8 & 4.9 & $16.8 \pm 0.7^\circ$ & $14.6 \pm 0.5^\circ$ & $(5.0 \pm 0.4)\times10^{-5}$ \\
$^{22}$Na  & 0, 0       & $0.5,\,2.0$    & 5.0 & 5.2 & $12.9 \pm 0.7^\circ$ & $12.4 \pm 0.4^\circ$ & $(7.8 \pm 0.6)\times10^{-6}$ \\
$^{22}$Na  & 0, 10      & $1.6,\,8.0$    & 5.1 & 4.8 & $13.4 \pm 0.8^\circ$ & $12.9 \pm 0.4^\circ$ & $(7.7 \pm 0.6)\times10^{-6}$ \\
$^{22}$Na  & 0, 30      & $1.7,\,22$     & 3.6 & 5.1 & $14.3 \pm 0.6^\circ$ & $14.1 \pm 0.5^\circ$ & $(7.8 \pm 0.6)\times10^{-6}$ \\
\hline
\end{tabular}%
}
\end{table}

\section{Discussion}
\label{discussion}


The results summarized in Table~\ref{tab:reco_exp_results} show that for the $^{137}$Cs point-source, the central and off-axis positions were reconstructed within 2--3~mm of their true locations, within the experimental uncertainty. The imaging of the $^{22}$Na point-source exhibited similar agreement, for source positions on, or close to detector axis. Larger deviation in the reconstructed position for the source at Y=30 mm off-axis, was investigated via Geant4 simulation of the setup. First, the simulated events, without detector effects, and without the strict kinematic cuts, were passed through the reconstruction algorithm, which yielded the source image at the correct position. Second, the detector effects (energy and position resolution) were implemented, with loose kinematic selection, which again resulted in the correct source image. Finally, when the strict kinematic selection $E_{scatter}>120$ keV and inter-pixel distance $d>13$ mm, were applied, the simulated image underestimated the off-axis shift by the same amount as in the experimental case. This strict selection required to suppress the noise at low energies, could be relaxed with dedicated low-noise electronics. 
According to Geant4 simulations \cite{dash2026thesis}, the loosening of the inter-pixel distance cut should remove the underestimation of the off-axis position, while the relaxation of the 120 keV threshold should improve the sensitivity.

Spatial resolutions ($\sigma_\mathrm{x}$ and $\sigma_\mathrm{y}$) are consistent between experiment and simulation, with $\sigma_\mathrm{x}=4.3$--6.5~mm and $\sigma_\mathrm{y}=5.2$--5.9~mm for $^{137}$Cs, and $\sigma_\mathrm{x}=2.7$--5.7~mm and $\sigma_\mathrm{y}=3.8$--5.2~mm for $^{22}$Na. ARM distributions similarly agree, confirming accurate Compton scattering reconstruction.  

The intrinsic efficiency for $^{22}\mathrm{Na}$ was approximately a factor of six lower than that for $^{137}\mathrm{Cs}$ ($\sim 7.7\times10^{-6}$). This difference is mainly attributed to the stricter event-selection conditions for $^{22}\mathrm{Na}$, particularly the narrower scatterer-energy range (120--140~keV compared with 120--160~keV for $^{137}\mathrm{Cs}$) and the larger minimum inter-pixel distance (13~mm compared with 10~mm), which reduce the number of usable Compton events.

Overall, experimental results validate the detector model and reconstruction algorithms, with small discrepancies in off-axis positions and intrinsic efficiency for $^{22}$Na explained by data selection criteria and reduced event statistics. These findings demonstrate Compton imaging capability for both 662~keV and 511~keV gamma sources.

\section{Conclusion}

We report the design, construction, and characterization of a compact single-plane readout Compton camera based on pixelated GAGG:Ce scintillators coupled to silicon photomultipliers (SiPMs). The detector employs a novel single-side readout configuration, where gamma-ray interactions in dual scintillator layers are read out from one end through a light guide, enabling a compact and cost-effective system suitable for portable gamma imaging applications.

A full 64-element detector array was assembled, consisting of an $8\times8$ matrix of $3\times3\times3$~mm$^3$ GAGG:Ce scintillators optically coupled to a $3\times3\times20$~mm$^3$ light guide per element and read out by a 64-channel SiPM array. Energy calibration of individual pixels achieved resolutions of $8.9\%\pm1.9\%$ (front layer) and $10.8\%\pm1.6\%$ (back layer). Strict Compton event selection, based on inter-pixel distance, energy deposition, and scattering constraints, ensured the reconstruction of valid events for imaging.

Point-source imaging experiments were performed using $^{137}$Cs (662 keV) and $^{22}$Na (511 keV) sources at multiple positions within the central field of view (up to $\pm17^\circ$ off-axis)
Images were reconstructed using both Simple Back-Projection (SBP) and LM-MLEM algorithms. Angular resolutions of 14.3°–16.8° for 662 keV and 12.4°–14.3° for 511 keV photons were achieved, with intrinsic detection efficiencies of $4.3\times10^{-5}$–$5.0\times10^{-5}$ for $^{137}$Cs and $7.7\times10^{-6}$–$7.8\times10^{-6}$ for $^{22}$Na. 

The developed single-plane readout Compton camera demonstrates reliable gamma-ray imaging with moderate spatial resolution, confirming the feasibility of compact, low-cost detectors for source localization in applications requiring portability and rapid deployment. 
The present geometry, employing 20-mm-long light guides, also provides a basis for scaling the detector to larger active areas by placing multiple detector matrices side-by-side while maintaining the same light-guide geometry. Increasing the active detector area is expected to primarily improve the detection sensitivity. Further investigation of larger detector configurations using multiple matrices and their optimization with Geant4 simulations will be an important direction for future development, complemented by studies with wider field-of-view and simultaneous multiple source detection to provide detailed performance characterization.

\section*{Acknowledgments}
This work was supported by the Croatian Science Foundation through the Early Career Development Program under Grant DOK-2020-01-4393; through research projects IP-2018-01-8570 and IP-2022-10-3878; and by the ``Research Cooperability'' Program of the Croatian Science Foundation, co-funded by the European Union from the European Social Fund under the Operational Programme Efficient Human Resources 2014--2020, Grant PZS-2019-02-5829.

\end{document}